\documentstyle[aps,prl,epsf,floats,multicol,amssymb,tighten]{revtex}
\begin{document}

\newenvironment{tab}[1]
{\begin{tabular}{|#1|}\hline}
{\hline\end{tabular}}
\newcommand{\HG}{\hat{G}}
\newcommand{\Ht}{\hat{t}}
\newcommand{\HI}{\hat{I}}
\newcommand{\Hg}{\hat{g}}
\newcommand{\LL}{\langle \langle}
\newcommand{\RR}{\rangle \rangle}
\newcommand{\nonb}{\nonumber}
\newcommand{\SOP}{superconducting order parameter }
\newcommand{\SOPbis}{superconducting order parameter}
\newcommand{\DOS}{density of states }

\newcommand{\fig}[2]{\epsfxsize=#1\epsfbox{#2}} \reversemarginpar 
\bibliographystyle{prsty}

\title{Proximity effect in multiterminal hybrid structures}
\author{
H. Jirari, R. M\'elin,
N. Stefanakis}
\address{
Centre de Recherches sur les Tr\`es basses
temp\'eratures (CRTBT)\thanks{U.P.R. 5001 du CNRS,
Laboratoire conventionn\'e avec l'Universit\'e Joseph Fourier}\\
BP 166X, 38042 Grenoble Cedex, France}

\maketitle

\begin{abstract}
We consider the proximity effect
in multiterminal ferromagnet~/ superconductor (FSF) hybrid structures
in which two or three electrodes are connected to a superconductor.
We show that two competing
effects take place in these systems:
(i) pair breaking effects due
to the response to the exchange field induced in
the superconductor; (ii) a reduction of
the \SOP at the interface that takes place already 
in NS junctions. We focus on this second effect
that
dominates if the thickness of the S layer
is small enough. We consider
several single-channel electrodes connected to the same site.
We calculate the \SOP and the local density of state (LDOS).
With two ferromagnetic electrodes connected
to a superconductor we find that the \SOP in the
ferromagnetic alignment is larger than the \SOP
in the antiferromagnetic alignment ($\Delta_{\rm F}
>\Delta_{\rm AF}$), in agreement with
[Eur. Phys. J. B {\bf 25}, 373 (2002)]. If a third
spin polarized electrode is connected to a
superconductor we find that $\Delta_{\rm F}-\Delta_{\rm AF}$
can change sign as the transparency of the third electrode
increases. This can be understood from the fact that
the \SOP is reduced if pair correlations
among the ferromagnetic electrodes increase.
If the two ferromagnetic electrodes are within a finite 
distance we find Friedel oscillations in the
Gorkov function but we still obtain
$\Delta_{\rm F} > \Delta_{\rm AF}$.
\end{abstract}

\widetext

\section{Introduction}

The manipulation of entangled states of electrons
in condensed matter devices has focussed
an important interest recently.
The ground state of a superconductor
is a condensate of Cooper pairs that form
singlet states. Entangled states
of electrons can thus be manipulated in
transport experiments by extracting
Cooper pairs out of a superconductor.
Several experiments using a superconductor as a source
of entangled states of electrons have been 
proposed recently. For instance it was
shown in Ref.~\cite{Loss} that entangled states
of electrons can be manipulated in a double
dot experiment. A quantum teleportation
experiment using three quantum dots
has been proposed recently~\cite{quantum-tele}.
Another possible experiment
has also been proposed in Ref.~\cite{Martin}
in which a ``beam splitter'' is connected
to a superconductor. In this
situation noise correlations
can reveal information about electronic
entanglement~\cite{Martin}.
Other possible experiments in which several
ferromagnetic electrodes are connected to
a superconductor have been investigated
theoretically in
Refs.~\cite{Feinberg,Falci,Melin,Melin-Feinberg,Apinyan}.

There is a rich physics occurring
at a single ferromagnet~/~superconductor (FS) interface.
For instance Andreev reflection
is suppressed if the spin
polarization of the ferromagnetic metal increases.
This is because the incoming electron and the reflected
hole belong to different spin bands.  As a consequence
Andreev reflection can occur only in the
channels having both a spin-up and a spin-down
Fermi surface~\cite{deJong}. This theoretical
prediction was well verified in
experiments~\cite{Soulen,Upadhyay} and it was 
shown that with high transparency interfaces
the suppression of Andreev reflection by
spin polarization can be used to probe
the Fermi surface polarization~\cite{Soulen}.
The results of the Andreev reflection experiments
compare well with another method based
on spin polarized tunneling in the presence
of Zeeman splitting~\cite{Tedrow-Meservey}.
Another phenomenon taking place at FS interfaces
is that the pair amplitude induced in a ferromagnetic
metal can oscillate in space. This gives the
possibility of fabricating $\pi$ junctions
in which the Josephson relation changes
sign~\cite{Fulde-Ferrel,Larkin,Clogston,Demler,Buzdin1,Ryazanov,Kontos}.
It is also well established that 
FS multilayers present oscillations of the superconducting
critical temperature as the thickness of the ferromagnetic
layer is varied~\cite{Buzdin2,exp1,exp2,exp3,exp4,exp5}.
Other new phenomena related to 
the proximity effect have also been investigated
in diffusive FS
heterostructures~\cite{Lawrence,Vasko,Giroud,Petrashov2,Filip,Giroud2}.

Multiterminal hybrid structures consist of systems in which
several spin polarized electrodes are connected to
a superconductor and are controlled by
crossed Andreev reflection
processes in which the spin-up and spin-down electrons
making the Cooper pair can tunnel in different
ferromagnetic electrodes. 
Several theoretical
predictions have been made. For instance
the current circulating in one of the
electrodes can be controlled by the voltage
applied on another electrode~\cite{Falci,Melin-Feinberg}.
The conductance can be described in terms
of a conductance matrix~\cite{Falci} that
can be calculated from Keldysh formalism~\cite{Melin-Feinberg}.
Other predictions concern the proximity effect
in FSF trilayers~\cite{Baladie}.
It was shown in Ref.~\cite{Apinyan} that 
the two ferromagnetic
electrodes of the FSF trilayer
are coupled by pair correlations
and that the \SOP in the ferromagnetic alignment
can be larger than the \SOP in the antiferromagnetic
alignment ($\Delta_{\rm F}>\Delta_{\rm AF}$).
One goal of our article is to find out the range of
validity of this result and to bridge the gap
with other predictions based on Usadel equations~\cite{Baladie}.
It was shown in the context of Usadel equations
that the critical temperature (and therefore the
superconducting order parameter) is larger in the
antiferromagnetic alignment ($\Delta_{\rm AF}>
\Delta_{\rm F}$)~\cite{Baladie}. We show that the ballistic
models that we consider can lead also to 
$\Delta_{\rm AF} > \Delta_{\rm F}$ if one takes into
account the existence of a finite exchange field in
the superconductor, in the spirit of Ref.~\cite{deGennes}.
There are thus two competing mechanisms for the proximity
effect in FSF trilayers. The mechanism based on the
exchange field involves the ``11'' and ``22''
diagonal elements of the
Green's function of the superconductor connected
to the ferromagnetic electrodes. The other mechanism
based on pair correlations involves the ``12''
extra diagonal element. To understand in detail the
mechanism based on pair correlations we investigate
a device in which three ferromagnetic electrodes are
coupled to a superconductor. As a simplifying assumption
we suppose that the three electrodes are coupled
to the same site in the superconductor. The 
fabrication of three contacts within a coherence length
is not possible with present time state of the art technology.
Nevertheless these systems with three electrodes provide
idealized situations that are useful for understanding
theoretically
how the value of the \SOP depends on pair correlations
induced in the ferromagnetic electrodes.

From the point of view of the method we use
two complementary approaches: (i) an analytical
evaluation
of the high energy
behavior of the Gorkov function; (ii) exact diagonalizations
of the Bogoliubov-de Gennes equations. The analytical
approach is based on the simplest ballistic models and
can be a useful guideline for understanding more realistic
systems in the future, such as multichannel systems or
diffusive systems.
An important difference between the analytical and
numerical approaches lies in the fact that
in the analytical
calculation we suppose that the band-width of the
superconductor is much larger than the
superconducting gap like in realistic systems
for which $\Delta/D \simeq 10^{-5}$.
As a consequence of this assumption
the integral appearing in the self-consistency
relation is dominated by the high frequencies.
By contrast
in the numerical simulation the band-width
of the superconductor is not small compared to
the superconducting gap (typically the ratio
between the bandwidth and the gap is $\Delta/D
\simeq 1/5$). The analytical calculation and
the numerical simulation correspond to different
regimes and this is why all the analytical
results cannot be confirmed by the numerical
simulation. The numerical simulation is useful
to discuss the behavior of the local density
of states (LDOS), the case of a partial
spin polarization and the spatial variation
of the pair amplitude.

The article is organized as follows.
Necessary technical preliminaries are given in
section~\ref{sec:prelim}.
The mechanism based on the exchange field induced in
the superconductor
is investigated in sections~\ref{sec:NF} and~\ref{sec:FS}.
The remaining of the article is devoted to the other
mechanism based on pair correlations.
In section~\ref{sec:FSF} we discuss the proximity
effect in FSF heterostructures and generalize
the results obtained in Ref.~\cite{Apinyan}.
Namely we show that the \SOP is
larger if the ferromagnetic electrodes
have a parallel spin orientation.
Multiterminal structures in which three
electrodes are connected to a superconductor
are discussed in section~\ref{sec:multiterminal}.
Friedel oscillations in the Gorkov function are
analyzed in section~\ref{sec:Friedel-Gorkov}.
Concluding remarks
are given in section~\ref{sec:conclusion}.

\section{Preliminaries}
\label{sec:prelim}

\subsection{Green's function formalism}
\label{sec:prelim-Green}
In this section we provide the technical details on Green's
function formalism that is used throughout the article.
\subsubsection{The models}
The superconductor is described by a
BCS lattice model
$$
{\cal H}_{\rm BCS} = \sum_{\langle \alpha , \beta
\rangle , \sigma} - t \left(
c_{\alpha,\sigma}^+ c_{\beta,\sigma}
+c_{\beta,\sigma}^+ c_{\alpha,\sigma} \right)
+ \sum_\alpha \left( \Delta_\alpha
c_{\alpha,\uparrow}^+ c_{\alpha,\downarrow}^+
+ \Delta_\alpha^* c_{\alpha,\downarrow}
c_{\alpha,\uparrow} \right)
,
$$
where $\Delta_\alpha$ is the pairing interaction.
The brackets indicate that hopping is 
between nearest neighbors.
The ferromagnetic electrodes are described by
a lattice Stoner model
$$
{\cal H}_{\rm Stoner} = \sum_{\langle i,j
\rangle,\sigma} -t \left( c_{i,\sigma}^+
c_{j,\sigma} + c_{j,\sigma}^+ 
c_{i,\sigma} \right)
-h \sum_i 
\left( c_{i,\uparrow}^+ c_{i,\uparrow}
-c_{i,\downarrow}^+ c_{i,\downarrow} \right)
,
$$
where $h$ is the exchange field.
An exchange field smaller than the bandwidth
corresponds to a partially polarized ferromagnet
and an exchange field larger than the bandwidth
corresponds to a half-metal ferromagnet in which
only majority spins are present.
We note $\rho^S_0$ the normal state 
density of states of the
superconductor and $\rho^F_\sigma$ the spin-$\sigma$
density of states in the ferromagnetic electrodes.
We use Greek symbols $\alpha$, $\beta$, $\gamma$
for the sites in the superconductor and Latin symbols
$a$, $b$, $c$ for the sites in the ferromagnetic electrodes.
The tunnel Hamiltonian coupling the superconductor and
the ferromagnetic electrodes takes the form
\begin{equation}
\label{eq:H-tunnel}
\hat{\cal W} = \sum_{k,\sigma} t_{a_k,\alpha_k}
\left( c_{a_k,\sigma}^+ c_{\alpha_k,\sigma}
+ c_{\alpha_k,\sigma}^+ c_{a_k,\sigma} \right)
,
\end{equation}
where the sum over $k$ runs over all contacts
between the superconductor and the ferromagnetic
electrodes.

\subsubsection{Green's functions}
The Green's functions of a connected system are obtained
by solving the Dyson equation in the Nambu representation:
\begin{equation}
\label{eq:Dyson}
\hat{G}^{R,A} = \hat{g}^{R,A}
+ \hat{g}^{R,A} \otimes \hat{\Sigma}
\otimes \hat{G}^{R,A}
,
\end{equation}
where the self-energy $\hat{\Sigma}$ contains
all couplings of the tunnel
Hamiltonian given by~(\ref{eq:H-tunnel}).
The Green's functions $g$ correspond to the 
``disconnected'' system in which 
$t_{a_k,\alpha_k}=0$ (see Eq.~\ref{eq:H-tunnel}).
The Dyson equation~(\ref{eq:Dyson}) is used
to calculate the Green's functions of the connected
system in which an electron in the superconductor
can make excursions in the ferromagnetic electrodes.

We use the following notation for the
Nambu representation of the advanced
and retarded propagators of the disconnected system:
$$
\hat{g}^{A,R} (t,t') = \left(
\begin{array}{cc}
g^{A,R}(t,t') & f^{A,R}(t,t') \\
f^{A,R}(t,t') & g^{A,R}(t,t') 
\end{array} \right)
,
$$
with
\begin{eqnarray}
g^A(t,t') &=& -i \theta(t-t')
\langle \left\{ c_{i,\uparrow}(t),
c_{j,\uparrow}^+(t') \right\} \rangle\\
f^A(t,t') &=& -i \theta(t-t')
\langle \left\{ c_{i,\uparrow}(t),
c_{j,\downarrow}(t') \right\} \rangle
,
\end{eqnarray}
and we use the following notation for the Nambu
representation of the density of states:
$$
\hat{\rho} = \left( \begin{array}{cc}
\rho_g & \rho_f \\
\rho_f & \rho_g \end{array} \right)
,
$$
with $\rho_g=\frac{1}{\pi} \mbox{Im}(g^A)$
and $\rho_f={1 \over \pi} \mbox{Im}(f^A)$.
The Keldysh Green's function is obtained
through the Dyson-Keldysh equation
\begin{equation}
\hat{G}^{+,-} = \left[ \hat{I}
+ \hat{G}^R \otimes \hat{\Sigma} \right]
\otimes \hat{g}^{+,-} \otimes
\left[ \hat{I} + \hat{\Sigma} 
\otimes \hat{G}^A \right]
,
\label{eq:Keldysh}
\end{equation}
where $\hat{g}_{i,j}^{+,-} = 
2 i \pi n_F(\omega-\mu_{i,j})
\hat{\rho}_{i,j}$.

\subsubsection{Superconducting order parameter
and transport properties}
In equilibrium the Keldysh Green's function
defined by Eq.~(\ref{eq:Keldysh}) simplifies
into
\begin{equation}
\label{eq:Keldysh-eq}
\hat{G}^{+,-}_{\rm eq} = 
n_F(\omega-\mu_0) \left( \hat{G}^A
-\hat{G}^R \right)
,
\end{equation}
where $\mu_0$ is the chemical potential,
identical in all conductors.
The form~(\ref{eq:Keldysh-eq}) of the
Keldysh Green's function at equilibrium
can be used to
obtain the \SOP in the superconductor
{\sl via} the self-consistency equation
\begin{equation}
\label{eq:self-con}
\Delta_\beta = - U \int \frac{d \omega}{2 i \pi}
G^{+,-,1,2}_{\beta,\beta}(\omega)
.
\end{equation}

\subsubsection{Green's function of an isolated normal metal}
Evaluating the spectral representation for a normal metal leads to
\begin{equation}
\label{eq:g-normal-metal}
g^A_{\alpha,\beta}(\omega) =
- \frac{m a_0^2}{\hbar^2} \frac{a_0}{2\pi R_{\alpha,\beta}}
exp{\left(-i k_F R -i \frac{\omega R_{\alpha,\beta}}
{v_F} \right)}
.
\end{equation}
The local density of states is given by
\begin{equation}
\label{eq:rho-loc-metal}
\label{eq:dos-metal}
\rho_{\rm loc}(\omega) = \frac{1}{\pi} \mbox{Im}
\left[ g^A(\omega) \right] = \frac{1}{2\pi^2} \frac{m a_0^3}
{\hbar^2} \left(k_F+\frac{\omega}{v_F}\right) 
.
\end{equation}

\subsubsection{Green's function of an isolated insulator}
We consider a semi-conductor model having
a valence band at energy $\epsilon(k)=-\Delta-\hbar^2 k^2/(2 m^*)$
and a conduction band at energy $\epsilon(k)
=\Delta+\hbar^2 h^2/(2m^*)$. This model behaves like
a band insulator if $|\omega|<\Delta$.
The Green's functions
in the energy range $|\omega|<\Delta$
are given by
\begin{equation}
\label{eq:green-iso}
g_{\alpha,\beta}(\omega) = \frac{m^* a_0^2}{\hbar^2}
\frac{a_0}{2\pi R_{\alpha,\beta}}
\left[ 
\exp{\left(-\frac{\sqrt{2m^* a_0^2}}{\hbar}\sqrt{\Delta+\omega} R
\right)} 
- \exp{\left(-\frac{\sqrt{2m^* a_0^2}}{\hbar} \sqrt{\Delta-\omega}R
\right)}
\right]
.
\end{equation}

\subsubsection{Green's functions of an isolated superconductor}
The Green's function of an isolated superconductor with a
uniform \SOP $\Delta_0$ is obtained
by evaluating the spectral representation~\cite{Melin-Feinberg}:
\begin{equation}
\label{eq:Green-below}
\hat{g}_{\alpha,\beta}(\omega) = \frac{2 m a_0^2}
{\hbar^2} \frac{a_0}{2\pi R}
\exp{\left[ - \frac{R_{\alpha,\beta}}{\xi(\omega)} \right]}
\left\{ \frac{\sin{\varphi_{\alpha,\beta}}}
{\sqrt{\Delta_0^2-\omega^2}} \left[ \begin{array}{cc}
-\omega & \Delta_0 \\ \Delta_0 & - \omega \end{array}
\right] + \cos{\varphi_{\alpha,\beta}}
\left[ \begin{array}{cc} - 1 & 0 \\ 0 & 1 \end{array} \right]
\right\}
.
\end{equation}
Above the superconducting gap the retarded Green's
function takes the form
\begin{equation}
\label{eq:Green-above}
\hat{g}^R_{\alpha,\beta} = \frac{2 m a_0^2}
{\hbar^2} \frac{a_0}{2\pi R}
\exp{\left[i \frac{\sqrt{\omega^2-\Delta_0^2}
R_{\alpha,\beta}}{v_F}\right]}
\left\{ \frac{i \sin{\varphi_{\alpha,\beta}}}{\sqrt{\omega^2-\Delta_0^2}}
\left[ \begin{array}{cc} - \omega & \Delta_0\\
\Delta_0 & - \omega \end{array} \right] 
+ \cos{\varphi_{\alpha,\beta}}
\left[ \begin{array}{cc} -1 & 0 \\ 0 & 1 \end{array} \right]
\right\}
.
\end{equation}
The phase variable in Eqs.~(\ref{eq:Green-below})
and~(\ref{eq:Green-above}) is $\varphi_{\alpha,\beta}
=k_F R_{\alpha,\beta}$.
The ``local'' propagators corresponding to
$\alpha=\beta$ are described by
$\varphi_{\alpha,\beta}=k_F R_0=\pi/2$~\cite{Cuevas}:
\begin{equation}
\label{eq:g-loc}
\hat{g}_{\rm loc}(\omega) = \frac{2 m a_0^2}{\hbar^2}
\frac{a_0}{2\pi R_0}
\frac{1}{\sqrt{\Delta_0^2-\omega^2}}
\left[ \begin{array}{cc}
-\omega & \Delta_0 \\
\Delta_0 & - \omega \end{array} \right]
.
\end{equation}

%%%%%%%%%%%%%%%%% FIGURE %%%%%%%%%%%%%%%%%%%%%%%%%
\begin{figure}[thb]
\centerline{\fig{4cm}{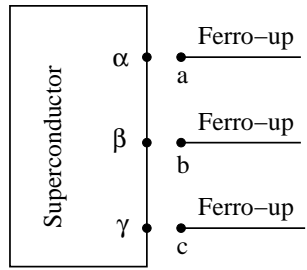}} 
\medskip
\caption{Schematic representation of the model
considered in section~\ref{sec:Dyson-matrix} in which
three half-metal ferromagnetic
electrodes are connected to a superconductor.
} 
\label{fig:simple}
\end{figure}
%%%%%%%%%%%%%%%%%%%%%%%%%%%%%%%%%%%%%%%%%%%%%%%%%%
\subsubsection{Dyson matrix}
\label{sec:Dyson-matrix}
In this section we provide a derivation of the Dyson matrix
in the simple case of three half-metal ferromagnetic
electrodes connected to a superconductor (see Fig.~\ref{fig:simple}).
If $\lambda$ is an arbitrary site in the superconductor, the
Dyson equation~(\ref{eq:Dyson}) becomes
\begin{equation}
\label{eq:DysonA}
\hat{G}^{a_n,\lambda} = \left[
\begin{array}{cc} K_{1,1}^{a_n,\lambda} &
K_{1,2}^{a_n,\lambda} \\
-K_{2,1}^{a_n,\lambda} & -K_{2,2}^{a_n,\lambda}
\end{array} \right]
+ \sum_m \left[ \begin{array}{cc}
K_{1,1}^{a_n,\alpha_m} & -K_{1,2}^{a_n,\alpha_m} \\
-K_{2,1}^{a_n,\alpha_m} & K_{2,2}^{a_n,\alpha_m}
\end{array} \right] t_{\alpha_m,a_m}
\hat{G}^{a_m,\lambda}
,
\end{equation}
with
$K_{i,j}^{a_n,\alpha_m} = g_{i,i}^{a_n,a_n}
t_{a_n,\alpha_n} g_{i,j}^{\alpha_n,\alpha_m}$,
and
\begin{equation}
\label{eq:DysonB}
\hat{G}^{a_n,\lambda} = \left[
\begin{array}{cc} G_{1,1}^{a_n,\lambda} &
G_{1,2}^{a_n,\lambda} \\
G_{2,1}^{a_n,\lambda} & G_{2,2}^{a_n,\lambda}
\end{array} \right]
.
\end{equation}
Eqs.~(\ref{eq:DysonA}) and~(\ref{eq:DysonB})
are valid for an arbitrary spin polarization.
In the case of the heterostructure on Fig.~\ref{fig:simple}
the explicit form of the Dyson matrix is the following:
$$
\left[ \begin{array}{ccc}
1-K_{1,1}^{a,\alpha}t_{\alpha,a} &
-K_{1,1}^{a,\beta} t_{\beta,b} &
-K_{1,1}^{a,\gamma} t_{\gamma,c}\\
-K_{1,1}^{b,\alpha} t_{\alpha,a} &
1-K_{1,1}^{b,\beta} t_{\beta,b} &
-K_{1,1}^{b,\gamma} t_{\gamma,c} \\
-K_{1,1}^{c,\alpha} t_{\alpha,a} &
-K_{1,1}^{c,\beta} t_{\beta,b} &
1-K_{1,1}^{c,\gamma} t_{\gamma,c} 
\end{array}
\right]
\left[ \begin{array}{c}
G_{1,1}^{a,\lambda} \\
G_{1,1}^{b,\lambda} \\
G_{1,1}^{c,\lambda} \end{array} \right]
=\left[ \begin{array}{c}
K_{1,1}^{a,\lambda}\\
K_{1,1}^{b,\lambda}\\
K_{1,1}^{c,\lambda} \end{array} \right]
.
$$
In sections~\ref{sec:FS} and~\ref{sec:FSF}
and in the Appendices
we invert a similar form of the Dyson matrix
in models involving four channels.
\subsection{Exact diagonalizations of the
Bogoliubov-de Gennes Hamiltonian}
\label{sec:prelim-BDG}

In the numerical simulations based on exact
diagonalizations of the Bogoliubov-de Gennes
Hamiltonian we use a lattice Hubbard model
\begin{equation}
\label{eq:H-Hubbard}
{\cal H}= \sum_{\langle \alpha,\beta \rangle,\sigma}
-t \left( c_{\alpha,\sigma}^+ c_{\beta,\sigma}
+ c_{\beta,\sigma}^+ c_{\alpha,\sigma} \right)
+ \mu \sum_{\alpha,\sigma} n_{\alpha,\sigma}
+ \sum_{\alpha,\sigma} \mu_\alpha^I
n_{\alpha,\sigma} + \sum_{\alpha,\sigma}
h_{\alpha,\sigma} n_{\alpha,\sigma}
+ V_0 \sum_\alpha n_{\alpha,\uparrow}
n_{\alpha,\downarrow}
.
\end{equation}
In Eq.~(\ref{eq:H-Hubbard}) $n_{\alpha,\sigma}
=c_{\alpha,\sigma}^+ c_{\alpha,\sigma}$
is the electron number operator at site $\alpha$,
$\mu$ is the chemical potential,
$h_{\alpha,\sigma}=-h \sigma_z$
is the exchange field in the ferromagnetic region
and $\sigma_z=\pm 1$ is the eigenvalue of the $z$
component of the Pauli matrix. $V_0$ is the
on-site interaction. We use negative values
of $V_0$ corresponding to attractive interaction.
To simulate
the effect of depletion of the carrier density
at the surface the site-dependent impurity potential
$\mu^I_\alpha$ is set to a sufficiently large value
at the surface sites. This prohibits electron
tunneling over these sites. Within a mean
field approximation Eq.~(\ref{eq:H-Hubbard})
reduces to the Bogoliubov-de Gennes
equations~\cite{deGennes-book,Stefanakis}:
\begin{equation}
\label{eq:Bogo1}
\left( \begin{array}{cc} \hat{\xi} & \hat{\Delta}\\
\hat{\Delta}^* & - \hat{\xi} \end{array} \right)
\left( \begin{array}{c} u_{n,\uparrow}(r_\alpha) \\
v_{n,\downarrow}(r_\alpha) \end{array} \right)
= \epsilon_{n,\gamma_1} \left( \begin{array}{c}
u_{n,\uparrow}(r_\alpha) \\ v_{n,\downarrow}(r_\alpha)
\end{array} \right),
\end{equation}
and
\begin{equation}
\label{eq:Bogo2}
\left( \begin{array}{cc} \hat{\xi} & \hat{\Delta}\\
\hat{\Delta}^* & - \hat{\xi} \end{array} \right)
\left( \begin{array}{c} u_{n,\downarrow}(r_\alpha) \\
v_{n,\uparrow}(r_\alpha) \end{array} \right)
= \epsilon_{n,\gamma_2} \left( \begin{array}{c}
u_{n,\downarrow}(r_\alpha) \\ v_{n,\uparrow}(r_\alpha)
\end{array} \right)
,
\end{equation}
such that
\begin{eqnarray}
\label{eq:xi}
\hat{\xi} u_{n,\sigma}(r_\alpha)&=&-t
\sum_{\hat{\delta}} u_{n,\sigma}(r_\alpha+\hat{\delta})
+ \left( \mu^I(r_\alpha) + \mu \right)
u_{n,\sigma}(r_\alpha)
+ h_\alpha \sigma_z u_{n,\sigma}(r_\alpha)\\
\hat{\Delta} u_{n,\sigma}(r_\alpha)&=&
\Delta_0(r_\alpha) u_{n,\sigma}(r_\alpha)
\label{eq:Delta}
,
\end{eqnarray}
and where the pair potential is defined by
$$
\Delta_0(r_\alpha)=V_0 \langle c_\uparrow(r_\alpha)
c_\downarrow(r_\alpha) \rangle
.
$$
In Eq.~(\ref{eq:xi}) $\hat{\delta}=\hat{x},-\hat{x},
\hat{y},-\hat{y}$ denotes the four directions of
the square lattice. $\hat{\delta}=\hat{z},
-\hat{z}$ corresponds to the two directions
in the ferromagnetic electrodes.
The self-consistency equation
takes the form
\begin{equation}
\label{eq:pair-ampli}
\Delta_0(r_\alpha)=\frac{V_0(r_\alpha)}{2} F(r_\alpha)
=\frac{V_0(r_\alpha)}{2} \sum_n \left[
u_{n,\uparrow}(r_\alpha) v_{v,\downarrow}^*(r_\alpha)
\tanh{\left( \frac{\beta \epsilon_{n,\gamma_1}}{2}
\right)}+u_{n,\downarrow}(r_\alpha)
v_{n,\uparrow}^*(r_\alpha) \tanh{\left(\frac{\beta
\epsilon_{n,\gamma_2}}{2} \right)} \right]
,
\end{equation}
where $\beta$ is the inverse temperature.
We start from an approximate solution for the
gap profile $\Delta_0(r_\alpha)$. After exact diagonalizations
of Eqs.~(\ref{eq:Bogo1}) and~(\ref{eq:Bogo2}) we obtain
$u_{n,\sigma}(r_\alpha)$ and 
$v_{n,\sigma}(r_\alpha)$. The quasiparticle amplitudes
are inserted into Eq.~(\ref{eq:pair-ampli})
and a new gap function is evaluated which is then inserted
into Eq.~(\ref{eq:Delta}) and we iterate until a sufficient
precision has been obtained. Although the pair
potential $\Delta_0(r_\alpha)$ is zero in the
ferromagnet the pair amplitude $F(r_\alpha)$ is
non zero. The LDOS at site
$\alpha$ is given by
\begin{equation}
\label{eq:LDOS-def}
\rho_\alpha(E) = - \sum_{n,\sigma}
\left[ |u_{n,\sigma}(r_\alpha)|^2 f'(E-\epsilon_n)
+ |v_{n,\sigma}(r_\alpha)|^2 f'(E+\epsilon_n) \right]
,
\end{equation}
where $f'$ is the derivative of the Fermi function
$f(\epsilon) = \frac{1}{1+\exp{(\epsilon/k_B T)}}$.

\section{Friedel oscillations at a single NF interface}
\label{sec:NF}
\begin{figure}[thb]
\centerline{\fig{8cm}{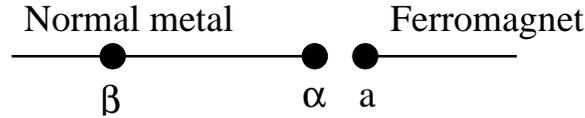}} 
\medskip
\caption{Schematic representation of the 
single-channel junction
between a normal metal and a ferromagnet. $R$ is the
distance between sites $\alpha$ and $\beta$.
} 
\label{fig:schem}
\end{figure}

We start by considering
a single interface between a normal metal and a spin
polarized ferromagnet.
We calculate the exchange field
at site $\beta$
in the normal metal, induced by the proximity of the
ferromagnet (see Fig.~\ref{fig:schem}).
%%%%%%%%%%%%%%%%% FIGURE %%%%%%%%%%%%%%%%%%%%%%%%%
%%%%%%%%%%%%%%%%%%%%%%%%%%%%%%%%%%%%%%%%%%%%%%%%%%
%%%%%%%%%%%%%%%%%%%%%%%%%%%%%%%%%%%%%%%%%%%%%%%%%%%%%%%%%%%%
\begin{figure}
\centerline{\fig{5cm}{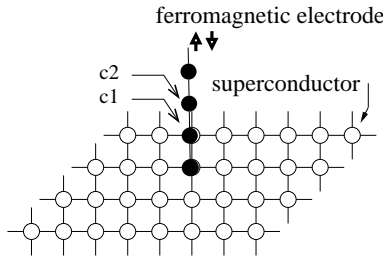}}
\medskip
\caption{The geometry used in the numerical simulations
in section~\ref{sec:FS}. A one-dimensional ferromagnetic
electrode is connected to a two-dimensional superconductor.
}
\label{fig:schema-1elec}
\end{figure}
%%%%%%%%%%%%%%%%%%%%%%%%%%%%%%%%%%%%%%%%%%%%%%%%%%%%%%%%%%%%
To order $t^2$ the Green's function at site $\beta$
is given by
\begin{equation}
\label{eq:G-beta}
G_{\beta,\beta} = g_{\beta,\beta} + 
g_{\beta,\alpha} t_{\alpha,a} g_{a,a} t_{a,\alpha}
g_{\alpha,\beta}
.
\end{equation}
$g_{\alpha,\beta}^A$ is given by Eq.~(\ref{eq:g-normal-metal})
and $g_{a,a}^A$ is decomposed in a real and imaginary
part:
$g_{a,a}^A=g_{a,a}^{(R)} + i g_{a,a}^{(I)}$.
If the exchange field is small compared to the typical
energy scales (being the energy band-width for a metal
or the charge gap for an insulator) then
the spin-up and spin-down Green's functions take the form
$$
g_{a,a}^{(\sigma)}(\omega)=g_{a,a}^{(0)}(\omega)
+\sigma h_{\rm ex} \frac{\partial g_{a,a}^{(0)}(\omega)}
{\partial \omega}
,
$$
where $g_{a,a}^{(0)}(\omega)$ is the Green's function
in the absence of an exchange field.
We deduce the existence of Friedel oscillations
in the spin polarization at the Fermi surface:
\begin{equation}
\label{eq:Friedel-ex}
P_{\beta} = \frac{\tilde{\rho}_{\beta,\beta}^{(\uparrow)}
- \tilde{\rho}_{\beta,\beta}^{(\downarrow)}}
{\tilde{\rho}_{\beta,\beta}^{(\uparrow)}
+ \tilde{\rho}_{\beta,\beta}^{(\downarrow)}}
= 2 \pi t_\alpha^2 \frac{1}{v_F}
\left( \frac{1}{2\pi R} \right)^2 h_{\rm ex}
\left[ -\sin{(2 k_F R)} \frac{\partial g_{a,a}^{(R)}}{\partial 
\omega}(0)
+\cos{(2 k_F R)} \frac{\partial g_{a,a}^{(I)}}{\partial \omega}(0)
\right]
.
\end{equation}
For an insulating ferromagnet we
obtain from Eq.~(\ref{eq:green-iso})
that
${\partial g_{a,a}^{(R)}}/{\partial 
\omega}(0)$ is negative which leads to a positive spin polarization
at site $\alpha$.
For a metallic ferromagnet we
obtain from Eq.~(\ref{eq:dos-metal})
that ${\partial g_{a,a}^{(I)}}/{\omega}(0)$ is positive.
Since we use $k_F R_0 = \pi/2$ we deduce that 
spin polarization at site $\alpha$ is negative.
As far as the magnitude of the effective exchange field
$h_{\rm eff}$ is concerned
we see that $h_{\rm eff}$ is of order
$\left(t_{a,\alpha}/\epsilon_F\right)^2 h_{\rm ex}$
in the metallic
case and of order $t_{a,\alpha}^2 {\Delta}^{-1/2}
\left( {m^* a_0^2}/{\hbar^2} \right)^{3/2} h_{\rm ex}$ in
the insulating model.

%%%%%%%%%%%%%%%%%%%%%%%%%%%%%%%%%%%%%%%%%%%%%%%%%%%%%%%%%%%%
\begin{figure}
\centerline{\fig{5cm}{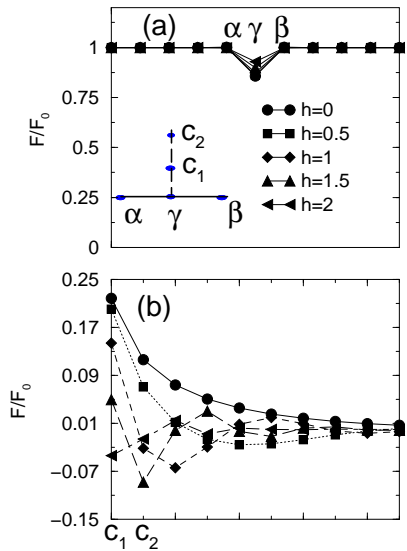}}
\medskip
\caption{
(a) Pair amplitude for different values of the exchange field
for several sites in the superconducting region
at the interface 
of one-dimensional ferromagnetic electrode and a two-dimensional 
superconducting system.
Sites $\alpha,\gamma,\beta$ 
belong to the superconductor while the sites $c_1,c_2$ 
belong to the ferromagnetic electrode.
(b) Pair amplitude for several sites in the ferromagnetic region
for different values of the exchange field.
The pair amplitude in the superconductor does not vary
much with the exchange field. The pair amplitude in the
ferromagnetic electrode shows an oscillatory behavior.
The pair amplitude decays monotonically in the
absence of spin polarization in the ferromagnetic
electrode.
}
\label{1dpa.fig}
\end{figure}
%%%%%%%%%%%%%%%%%%%%%%%%%%%%%%%%%%%%%%%%%%%%%%%%%%%%%%%%%%%%

%%%%%%%%%%%%%%%%%%%%%%%%%%%%%%%%%%%%%%%%%%%%%%%%%%%%%%%%%%%%
\begin{figure}
\centerline{\fig{5cm}{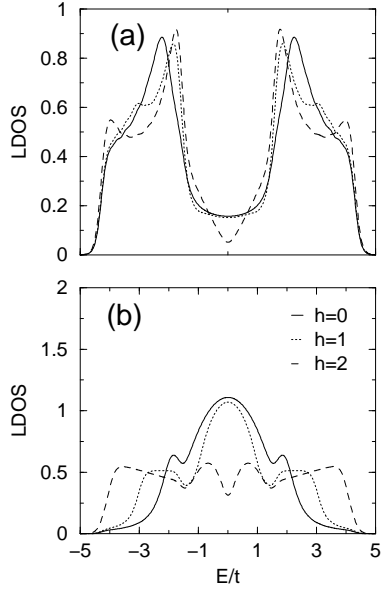}}
\medskip
\caption{
(a) The LDOS at site $\gamma$ in the superconductor
for different values of the exchange field, for the geometry 
of Fig.~\ref{1dpa.fig}.
(b) The LDOS at site $c_1$ in the ferromagnetic electrode
for different values of the exchange field.
The subgap LDOS is due to Andreev bound states.
}
\label{1ddos.fig}
\end{figure}
%%%%%%%%%%%%%%%%%%%%%%%%%%%%%%%%%%%%%%%%%%%%%%%%%%%%%%%%%%%%

We note that in the insulating case
the limit $R \rightarrow 0$ is well-defined 
and the value of spin polarization at site $\alpha$
does not depend on the value of the ultraviolet
cut-off $R_0$ introduced at short length scale.
In the metallic model the value of
the spin polarization is diverging if one takes the
limit $R \rightarrow 0$. Therefore 
the sign of spin polarization at site $\alpha$ depends 
on the value of the short distance cut-off $R_0$.
In the analytical approaches we use
a regularization in which we introduce a cut-off equal to
$k_F R_0 = \pi/2$ that is chosen similarly to Ref.~\cite{Cuevas}.
The justification of this choice of $R_0$ is that it leads
to a physically acceptable value of the local density of states.
The value of the density of states at the Fermi
energy for $R=R_0$ is given by
$$
\rho_{a,a}(R_0) = \frac{m a_0^2}{\hbar^2} 
\frac{a_0}{2\pi R_0} \sin{(k_F R_0)}.
$$
With the choice $k_F R_0=\pi/2$ of the short distance cut-off
the local density of state 
is equal to $\rho_{a,a}(R_0)=m a_0^3 k_F /
(\pi^2 \hbar^2)$, a positive value that is not far from
the exact density of states obtained for
$R_0=0$ (see Eq.~(\ref{eq:rho-loc-metal})).

\section{Proximity effect at a single FS interface}
\label{sec:FS}

\subsection{Local density of states}

We first use exact diagonalizations of the
Bogoliubov-de Gennes
equations (see section~\ref{sec:prelim-BDG})
to describe the proximity effect
at a single FS interface. This constitutes
a test of the numerical method and we
recover some of the results for the LDOS
obtained recently by means
of a recursion method (see Ref.~\cite{recursion}).

We consider a two-dimensional superconducting system
of $30 \times 30$ sites and we suppose fixed 
boundary conditions by setting the impurity potential
$\mu^I=100 t$ at the surface. The temperature
is $k_B T=0.1 t$ and the local attractive interaction in
the superconducting region is $V_0=-3.5 t$. On top
of the superconductor is attached a one-dimensional
superconducting electrode of $50$ sites (see
Fig.~\ref{fig:schema-1elec}).
The transparency of the interface is controlled
by changing the hopping element $t_c$
connecting sites on both sides of the interface
and we restrict here to the case where the transparency
of the interface is the same as inside the superconductor.
In this case the pair amplitude in the superconducting
system is not really modified as seen in Fig.~\ref{1dpa.fig}(a)
while in the ferromagnetic region the pair amplitude
oscillates around zero and the period of
oscillations decreases with increasing the exchange field
(see Fig.~\ref{1dpa.fig}(b)). In the case of a zero
exchange field the pair amplitude is decaying 
monotonically in the normal metal.

Due to the proximity effect the LDOS
shows a gap structure even for the sites within the ferromagnet
(see Fig.~\ref{1ddos.fig}(b)). The conductance peaks within the 
gap are due to Andreev bound states~\cite{deGennes-SaintJames}
and have been discussed recently for a three dimensional
FS interface using a recursion method~\cite{recursion}.
The residual values of the LDOS are reduced by the increase
of the exchange field. The Andreev bound states move
towards the Fermi level and cross the Fermi level
with increasing the exchange field.

\subsection{Exchange field in the superconductor}
From the discussion in section~\ref{sec:NF}
we deduce that a finite exchange field is induced in the
superconductor~\cite{deGennes}. The effect was also
found in recent numerical simulations in Ref.~\cite{Valls}.
In the metallic case and in the tunnel limit
the magnitude of the exchange
field is of order
$(t_{a,\alpha}/\epsilon_F)^2 h_{\rm ex}$.

An exchange field in
a superconductor is a pair-breaking perturbation
that tends to dissociate Cooper
pairs~\cite{deGennes-book,Tinkham}.
In this situation we expect that Cooper pairs couple
to the exchange field averaged over a length scale
equal to the BCS coherence length. A qualitative
argument can be made from the NF model in
section~\ref{sec:NF}.
Since the discussion is similar in
the two cases of metallic and insulating ferromagnets
we consider a metallic ferromagnet model only.
We obtain 
$$
\tilde{\rho}_{\beta,\beta}^{(\uparrow)}(\omega) -
\tilde{\rho}_{\beta,\beta}^{(\downarrow)} (\omega)
= \frac{t_{a,\alpha}^2 h_{\rm ex}}{4\pi^4 k_F a_0}
\left( \frac{m a_0^2}{\hbar^2} \right)^4
\left(\frac{a_0}{R}\right)^2
\cos{ \left[ 3 ( k_F + \omega/v_F) \right]}
,
$$
from what we deduce
$$
\int_{R_0}^{+\infty} 4 \pi R^2 dR
\int_0^D 
\left[ \tilde{\rho}_{\beta,\beta}^{(\uparrow)}(\omega) -
\tilde{\rho}_{\beta,\beta}^{(\downarrow)} (\omega) \right]
d\omega =-
\frac{t_{a,\alpha}^2 h_{\rm ex}}{3 \pi^3}
\left( \frac{m a_0^2}{\hbar^2} \right)^3
\int_{3 k_F R_0}^{3 \left(k_F+\frac{D}{v_F}\right)R_0}
\frac{\sin{u}}{u} du
.
$$
With $k_F R_0=\pi/2$
the spin polarization induced at long distance
in the superconductor
is positive but the local
spin polarization at the contact is negative.

The reduction of the superconducting order parameter
due to pair-breaking effects was
observed in FSF trilayers with insulating ferromagnets
in Refs.~\cite{Deutscher,Hauser}.
The effect is not visible in the point
contact experiments in Ref.~\cite{Soulen} even though
one might expect a reduction of the superconducting 
gap at the interface that would increase with increasing
the exchange field. It is likely that the 
point contact geometry used in Ref.~\cite{Soulen} plays
a role and that the effect should be more
pronounced for extended contacts.

%%%%%%%%%%%%%%%%% FIGURE %%%%%%%%%%%%%%%%%%%%%%%%%
\begin{figure}[thb]
\centerline{\fig{8cm}{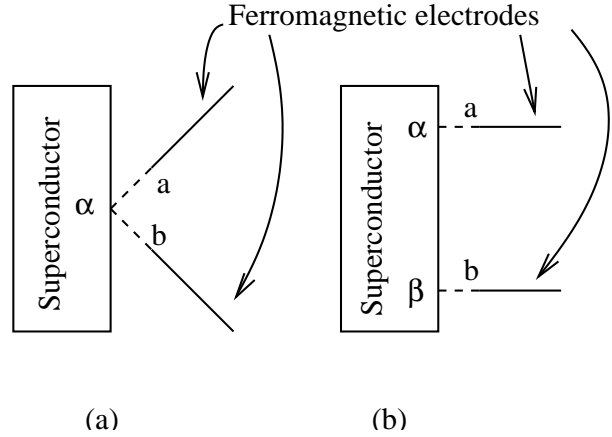}} 
\medskip
\caption{The two models considered in sections~\ref{sec:local-FSF}
and~\ref{sec:Friedel-Gorkov}. In (a) the two ferromagnetic
electrodes are connected to the same site in the
superconductor. In (b) the two ferromagnetic electrodes
are connected to two different sites in the superconductor.
The two ferromagnetic electrodes contain
a spin-up and a spin-down conduction band. The models
are thus solved in a $4 \times 4$ formalism.
} 
\label{fig:schema1}
\end{figure}
%%%%%%%%%%%%%%%%%%%%%%%%%%%%%%%%%%%%%%%%%%%%%%%%%%

%%%%%%%%%%%%%%%%% FIGURE %%%%%%%%%%%%%%%%%%%%%%%%%
\begin{figure}[thb]
\centerline{\fig{8cm}{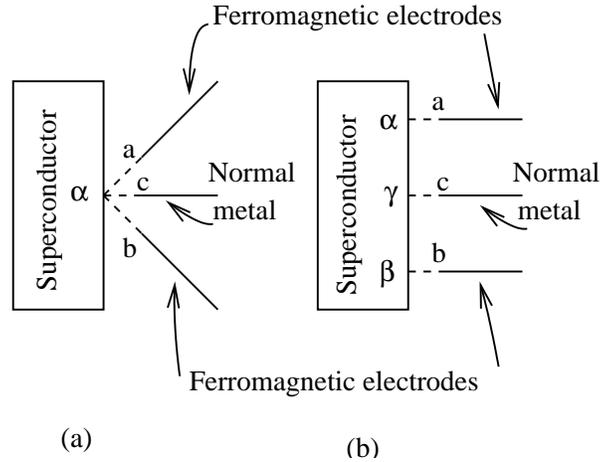}} 
\medskip
\caption{The model considered in
section~\ref{sec:multiterminal}. In (a) the three
electrodes are connected to the same site in the
superconductor. In (b) the three electrodes
are connected to two different sites in the superconductor.
We use the model (a) to calculate the \SOP.
} 
\label{fig:schema2}
\end{figure}
%%%%%%%%%%%%%%%%%%%%%%%%%%%%%%%%%%%%%%%%%%%%%%%%%%

\section{Proximity effect in FSF heterostructures: local models}
\label{sec:FSF}
Now we reconsider the proximity effect in a FSF heterostructure
in which two ferromagnetic electrodes are connected
to a superconductor. In this section as well as in
section~\ref{sec:multiterminal} we consider ``local'' models
in which two or three ferromagnetic electrodes are connected
to the same site in the superconductor
(see Figs.~\ref{fig:schema1}-(a) and~\ref{fig:schema2}-(a)).

%%%%%%%%%%%%%%%%% FIGURE %%%%%%%%%%%%%%%%%%%%%%%%%
\begin{figure}[thb]
\centerline{\fig{8cm}{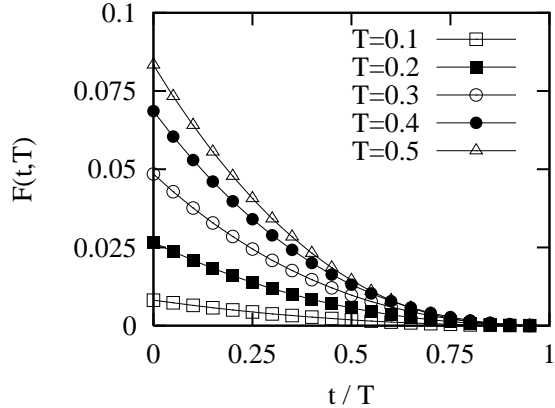}} 
\medskip
\caption{Variation of $F(t,T)$ for different
values of $T<1/2$. We obtain $F(t,T)>0$ which
shows that the ferromagnetic \SOP is larger
than the antiferromagnetic \SOP for
large interface transparencies.
The microscopic interaction $U$ is such that
the bulk superconducting gap is
$\Delta_{\rm bulk}=1$ and the bandwidth of
the superconductor is $D=100$.
} 
\label{fig:non-pert}
\end{figure}
%%%%%%%%%%%%%%%%%%%%%%%%%%%%%%%%%%%%%%%%%%%%%%%%%%

\subsection{Motivation}
\label{sec:physical}
\label{sec:mot}
Two effects can
play a role in the determination of the self-consistent
order parameter: 
\begin{itemize}
\item[(i)] {\sl The proximity effect} that takes place already
at a single NS interface. 
A pair amplitude is generated in the normal
metal and there is a reduction of the order parameter
on the superconducting side of the
interface~\cite{deGennes-book,Tinkham}.
For the models on Fig.~\ref{fig:schema1} the two
electrodes ending at sites ``a'' and ``b'' are coupled
by pair correlations only in the antiferromagnetic
alignment. The case of an antiparallel spin orientation
is thus qualitatively similar to a NS interface in which
case the order parameter in the superconductor is reduced
at the interface.
On the other no pair correlations are generated among
the ferromagnetic electrodes in the parallel alignment
if we consider half-metal ferromagnets. The case of parallel
magnetization is thus similar to an isolated
superconductor, from what we deduce that the \SOP
is larger in the ferromagnetic alignment.

\item[(ii)] {\sl Pair breaking effects} due 
the finite exchange field in the superconductor.
The exchange field is a pair-breaking perturbation
that tends to reduce the superconducting
order parameter. 
\end{itemize}

A relevant question is to determine which effect
would control real experiments on metallic FSF trilayers.
It is likely that the answer depends on the
geometry of the devices. 
For small interface transparencies
the exchange field in the superconductor
is of order $(t/ \epsilon_F)^2$ (see section~\ref{sec:FS}).
For a thin film in a parallel field the pair breaking
parameter is equal to $\alpha = D e^2 H^2 d^2/(6 \hbar c^2)$
with $D$ the diffusion constant, $H$ the magnetic field in
Tesla, $d$ the width of the superconductor. The 
difference between the critical temperatures in the
ferromagnetic and antiferromagnetic alignments
is given by
$k_B \Delta T_c = \pi \alpha/4$~\cite{Tinkham}
which is proportional to $t^4$ because $H$ is
proportional to $t^2$.
On the other hand if the physics is dominated by the 
proximity effect we deduce from Ref.~\cite{Melin}
that the difference between the superconducting order
parameters in the ferromagnetic and antiferromagnetic alignments
is of order
$\Delta_{\rm F}-\Delta_{\rm AF}= t^4 \Delta^{(0)}/(U \epsilon_F^3)$,
with $\Delta^{(0)}=D \exp{[-1/(U \rho_N)]}$ the BCS 
superconducting order parameter. We see that $\Delta T_c$
due to pair-breaking effects is proportional to $d^2$ and
we deduce that pair-breaking effects are
reduced if the width of the superconducting layer is reduced.

In the following we focus on the effect of pair correlations
induced in the ferromagnetic electrodes
and we
take for granted that the width of the superconductor
is small enough so that pair-breaking effects can be neglected.
We make the
further simplifying assumption of considering that all
ferromagnetic electrodes are connected to the same site
and we calculate the \SOP at this site. 
The case of two electrodes at a finite distance will be
considered in section~\ref{sec:Friedel-Gorkov}.
The qualitative
physics occurring in the schematic model with all electrodes
connected to the same site
can be described the
following rule: {\sl increasing
pair correlations in the ferromagnetic electrodes tends to
reduce the value of the self-consistent order
parameter in the superconductor}.

To illustrate this
let us first consider the case of two ferromagnetic
electrodes connected to a superconductor (see
Fig.~\ref{fig:schema1}-(a)).
Pair correlations among the two ferromagnets
are stronger in the antiferromagnetic alignment than
in the ferromagnetic alignment. As a consequence of
the rule given above the superconducting
order parameter is larger in the ferromagnetic alignment.
To make a more refined test of this rule 
we consider in
section~\ref{sec:multiterminal} a system in which three
electrodes are connected to a superconductor
(see Fig.~\ref{fig:schema2}-(a)). For simplicity we consider
that the three ferromagnetic electrodes are half-metal
ferromagnets. Electrode $c$ is supposed to have a 
spin-down orientation. Electrode $a$ has a spin-up orientation.
Electrode $b$ can have a spin-up orientation (ferromagnetic
alignment) or a spin-down orientation (antiferromagnetic
alignment).
If we consider first that $t_{c,\gamma}$ is small
we see that pair correlations are formed mainly
among electrodes
$a$ and $b$ if electrodes $a$ and $b$ have an opposite
spin orientation. We deduce that $\Delta_{\rm F}
>\Delta_{\rm AF}$ if $t_{c,\gamma}$ is small.
We consider now that $t_{c,\gamma}$ is
large and electrode $c$ has a spin-down orientation.
If electrodes $a$ and $b$ have a spin-up orientation
we see that pair correlations are formed mainly among
electrode $a$ and $c$ and among electrodes $b$ and $c$.
If electrode $a$ has a spin-up orientation and electrode
$b$ has a spin-down orientation we see that pair
correlations are formed mainly among electrodes $a$
and $c$. As a consequence the proximity effect is
stronger if electrodes $a$ and $b$ have a parallel
orientation. We deduce from the rule given above 
that $\Delta_{\rm F}<\Delta_{\rm AF}$ if $t_{c,\gamma}$
is large.

%%%%%%%%%%%%%%%%% FIGURE %%%%%%%%%%%%%%%%%%%%%%%%%
\begin{figure}[thb]
\centerline{\fig{6cm}{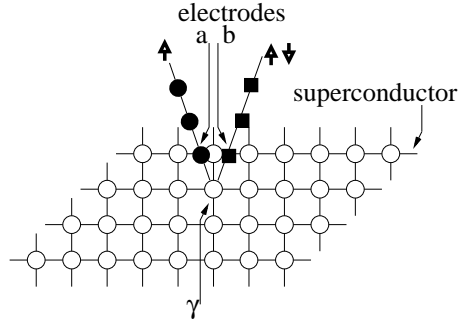}} 
\medskip
\caption{The geometry used in the numerical simulations
of the Bogoliubov-de Gennes Hamiltonian. Two one-dimensional
ferromagnetic electrodes are connected to the same site
of a two-dimensional superconductor. Electrode 
ending at site ``a''
is represented by filled circles and electrode
ending at site ``b''
is represented by filled squares.
} 
\label{fig:geom-2elec}
\end{figure}
%%%%%%%%%%%%%%%%%%%%%%%%%%%%%%%%%%%%%%%%%%%%%%%%%%

%%%%%%%%%%%%%%%%% FIGURE %%%%%%%%%%%%%%%%%%%%%%%%%
\begin{figure}[thb]
\centerline{\fig{6cm}{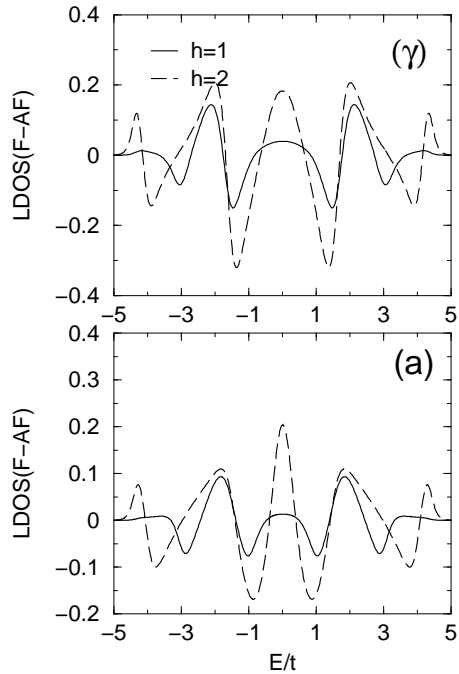}} 
\medskip
\caption{Difference between the LDOS
in the ferromagnetic
and antiferromagnetic spin alignments for the geometry
on Fig.~\ref{fig:geom-2elec}. The LDOS is represented
as a function of $E/t$,  
for two values of the exchange field
($h=1$ and $h=2$). The upper panel corresponds to
the LDOS at site $\gamma$ and the lower panel 
corresponds to the LDOS at site $a$
(see Fig.~\ref{fig:geom-2elec}). 
The subgap LDOS is larger in the ferromagnetic
alignment.
} 
\label{fig:dos-diff}
\end{figure}
%%%%%%%%%%%%%%%%%%%%%%%%%%%%%%%%%%%%%%%%%%%%%%%%%%

%%%%%%%%%%%%%%%%% FIGURE %%%%%%%%%%%%%%%%%%%%%%%%%
\begin{figure}[thb]
\centerline{\fig{6cm}{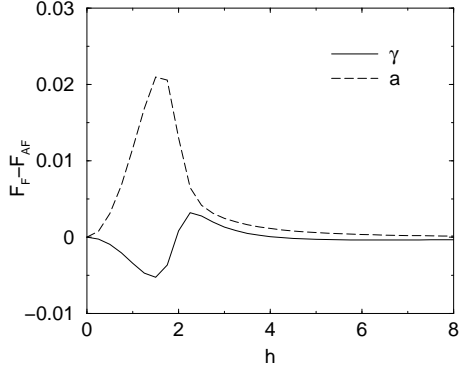}} 
\medskip
\caption{The difference between the pair amplitudes
in the ferromagnetic and antiferromagnetic alignments
for two electrodes connected to the same
site $\gamma$.
} 
\label{2ddif.fig}
\end{figure}
%%%%%%%%%%%%%%%%%%%%%%%%%%%%%%%%%%%%%%%%%%%%%%%%%%

\subsection{Two ferromagnetic electrodes connected
to the same site}
\label{sec:local-FSF}

\subsubsection{Sign of $\Delta_{\rm F}-\Delta_{\rm AF}$}
Let us start with the situation where two ferromagnetic
electrodes are connected to the same site (see
Fig.~\ref{fig:schema1}-(a)). We suppose
that each ferromagnetic electrode is partially 
spin polarized and
contains two spin
channels. The model is thus solved in a 
$4 \times 4$ formalism. The Green's function
$G^{A,1,2}_{\alpha,\alpha}$ is found to be
\begin{eqnarray}
\label{eq:Gorkov-local1}
G_{\alpha,\alpha}^{A,1,2} &=&
-i \pi \rho^S_0 {\Delta \over \omega} - i \pi \rho^S_0
{\Delta \over \omega} {1 \over {\cal D}}
\left \{ -f(x_{a,\uparrow}) - f(x_{a,\downarrow})
-f(x_{b,\uparrow}) - f(x_{b,\downarrow})
+2 f(x_{a,\uparrow}) f(x_{b,\uparrow})
+2 f(x_{a,\downarrow}) f(x_{b,\downarrow})\right.\\
\nonb
&+& f(x_{a,\uparrow}) f(x_{a,\downarrow})
+ f(x_{a,\uparrow})f(x_{b,\downarrow})
+ f(x_{a,\downarrow})f(x_{b,\uparrow})
+ f(x_{b,\uparrow})f(x_{b,\downarrow})
- f(x_{a,\uparrow})f(x_{a,\downarrow})f(x_{b,\downarrow})\\
\nonb
&-& \left.  f(x_{a,\uparrow})f(x_{a,\downarrow})f(x_{b,\uparrow})
- f(x_{a,\downarrow})f(x_{b,\uparrow})f(x_{b,\downarrow})
- f(x_{a,\uparrow})f(x_{b,\uparrow})f(x_{b,\downarrow})
- f(x_{a,\uparrow})f(x_{a,\downarrow})
f(x_{b,\uparrow})f(x_{b,\downarrow}) \right\}
,
\end{eqnarray}
with 
\begin{equation}
\label{eq:D-local1}
{\cal D} = \left( 1 - f(x_{a,\downarrow}) f(x_{b,\downarrow})
\right)
\left(1
- f(x_{a,\uparrow})f(x_{b,\uparrow}) \right)
.
\end{equation}
The interface transparencies
are parametrized by $f(x_i)=x_i/(1+x_i)$, with
$x_{a,\uparrow}=\pi^2 t_{a,\alpha}^2 \rho^F_{a,\uparrow} \rho^S_0$,
$x_{a,\downarrow}=\pi^2 t_{a,\alpha}^2 \rho^F_{a,\downarrow} \rho^S_0$,
$x_{b,\uparrow}=\pi^2 t_{b,\beta}^2 \rho^F_{b,\uparrow} \rho^S_0$,
$x_{b,\downarrow}=\pi^2 t_{b,\beta}^2 \rho^F_{b,\downarrow} \rho^S_0$.
The values of $f(x_i)$ are such that $0<f(x_i)<1/2$.
We do not consider the regime $1/2<f(x_i)<1$ for
which we find that the self-consistency relation
is unstable. 

We suppose that the two ferromagnets have an identical
density of states and
that the two contacts have identical transparencies.
In the ferromagnetic alignment we use the notation
$ f(x_{a,\uparrow})=f(x_{b,\uparrow})=T$ and
$ f(x_{a,\downarrow})=f(x_{b,\downarrow})=t$.
In the antiferromagnetic alignment we use the notation
$ f(x_{a,\uparrow})=f(x_{b,\downarrow})=T$ and
$ f(x_{a,\downarrow})=f(x_{b,\uparrow})=t$.
To second order in $T$ and $t$ we obtain
\begin{equation}
\label{eq:diffGalphaalpha}
G_{\alpha,\alpha}^{1,2,F} -
G_{\alpha,\alpha}^{1,2,AF} =
-i \pi \rho^S_0 {\Delta \over \omega}
(T-t)^2
.
\end{equation}
As a consequence in the regime of small interfaces transparencies
we obtain $\Delta_{\rm F} > \Delta_{\rm AF}$.
The regime of a large interface transparencies
can be treated by evaluating numerically the
difference between the Gorkov functions
$G_{\alpha,\alpha}^{1,2,F} -
G_{\alpha,\alpha}^{1,2,AF} = -i \pi \rho^S_0 {\Delta \over \omega}
F(T,t)$.
It is visible on Fig.~\ref{fig:non-pert}
that $F(T,t)$ is positive for any value of $T<1/2$ and $t<T$
so that the ferromagnetic \SOP is larger than
the antiferromagnetic \SOPbis, in agreement with
Ref.~\cite{Apinyan}.

\subsection{Local density of states and pair amplitude}
We calculated the LDOS for a partial spin polarization
by means
of exact diagonalizations of the Bogoliubov-de Gennes
Hamiltonian. The geometry of the simulation is represented
on Fig.~\ref{fig:geom-2elec}.
It is visible on Fig.~\ref{fig:dos-diff}
that
the low energy LDOS is larger
in the ferromagnetic alignment than in the
antiferromagnetic alignment. 

The variation of the difference $F_{\rm F}-F_{\rm AF}$
between the pair
amplitudes in the ferromagnetic and antiferromagnetic
alignments as a function of the
exchange field is shown on Fig.~\ref{2ddif.fig}.
The numerical simulation coincides
with the analytical model for a large exchange field
in the sense that the pair amplitude is larger in the
ferromagnetic alignment. However for
small values of the exchange field
we obtain $F_{\rm F} > F_{\rm AF}$ or 
$F_{\rm F} < F_{\rm AF}$ depending
on which lattice site is considered. 
The possible origin of the discrepancies between
the numerical simulation and the analytical model
are analyzed in the concluding section.

\section{Proximity effect in multiterminal hybrid structures:
local models}
\label{sec:multiterminal}
In this section we consider an heterostructure in which three
electrodes are connected at the same
site to a superconductor.
Electrodes
$a$ and $b$ on Fig.~\ref{fig:schema2} will
be considered to be half-metal
ferromagnets so that the models are solved
in a $4 \times 4$ formalism. Using partially
polarized ferromagnets would require to solve
a $6 \times 6$ formalism for which we could not carry
out the analytical calculation in the regime of
large interface transparencies.
The case of partially polarized ferromagnets will be
treated numerically within exact diagonalizations of the
Bogoliubov de Gennes equations.

\subsection{Three electrodes connected at the same site}
\label{sec:3elec-Delta}
We consider a model in which three electrodes
are connected to the same site $\alpha$
(see Fig.~\ref{fig:schema2}-(a)). 
We determine the variation of $\Delta_{\rm F}
-\Delta_{\rm AF}$ as a function of the transparency
$t_c$ of the contact with electrode $c$.
We show that
there is no change in the sign of $\Delta_{\rm F}
-\Delta_{\rm AF}$ as $t_c$ increases
if electrode $c$ is a normal
metal. There is a change of sign in
$\Delta_{\rm F} - \Delta_{\rm AF}$
as $t_c$ increases if electrode
$c$ is a ferromagnet with a spin-down
orientation.
This is in agreement with the qualitative discussion
in section~\ref{sec:mot}.

\subsection{Sign of $\Delta_{\rm F}
-\Delta_{\rm AF}$}
\label{sec:local-4channel}
\label{sec:sign-local}
\begin{figure}[thb]
\centerline{\fig{8cm}{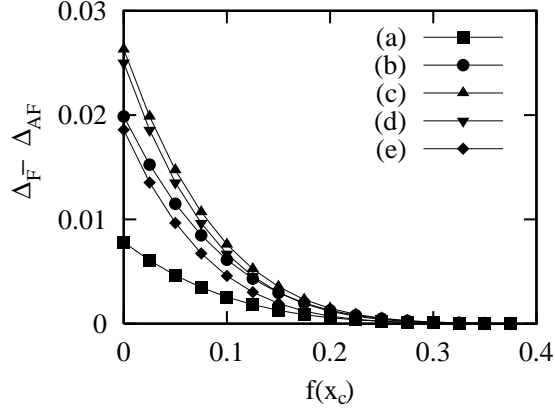}} 
\medskip
\caption{Variation of $\Delta^F_{\alpha,\alpha}
-\Delta^{AF}_{\alpha,\alpha}$ as a function of
the transparency of the contact with the
normal electrode $f(x_{c,\uparrow})
= f(x_{c,\downarrow})$ for different
values of the transparency of the contacts
with the ferromagnets $f(x_a)=f(x_b)$.
The different curves correspond to
(a) $f(x_a)=f(x_b)=0.05$, (b)
$f(x_a)=f(x_b)=0.1$, (c) $f(x_a)=f(x_b)=0.15$,
(d) $f(x_a)=f(x_b)=0.2$, (e) $f(x_a)=f(x_b)=0.25$.
We obtain
$\Delta_{\rm F} > \Delta_{\rm AF}$. The microscopic
interaction $U$ is such that the \SOP
of the isolated superconductor is
$\Delta_{\rm bulk}=1$ and the bandwidth of
the superconductor is $D=100$.
} 
\label{fig:appli4}
\end{figure}

In the ferromagnetic alignment the inversion
of the $4 \times 4$ Dyson matrix leads to
\begin{eqnarray}
G_{\alpha,\alpha}^{1,2} &=& -i \pi \rho^S_0
\frac{\Delta}{\omega}
-i \pi \rho^S_0 \frac{\Delta}{\omega}
\frac{1}{{\cal D}_{\rm F}}
\left\{ -f(x_a) -f(x_b) -f(x_{c,\uparrow})
-f(x_{c,\downarrow}) 
+2 f(x_a) f(x_b)
+2 f(x_a) f(x_{c,\uparrow})\right.\\
\nonb
&+& 2 f(x_b) f(x_{c,\uparrow})
+ f(x_a) f(x_{c,\downarrow})
+ f(x_b) f(x_{c,\downarrow})
+ f(x_{c,\uparrow}) f(x_{c,\downarrow})
-3 f(x_a) f(x_b) f(x_{c,\uparrow})
-f(x_a) f(x_b) f(x_{c,\downarrow})\\
\nonb
&-&\left. f(x_a) f(x_{c,\uparrow}) f(x_{c,\downarrow})
- f(x_b) f(x_{c,\uparrow}) f(x_{c,\downarrow})
+ f(x_a) f(x_b) f(x_{c,\uparrow}) f(x_{c,\downarrow})
\right\}
,
\end{eqnarray}
with
$$
{\cal D}_{\rm F} = 1-f(x_a) f(x_b)
-f(x_a) f(x_{c,\uparrow})
-f(x_b) f(x_{c,\uparrow})
+2 f(x_a) f(x_b) f(x_{c,\uparrow})
.
$$
In the antiferromagnetic alignment we find
\begin{eqnarray}
G_{\alpha,\alpha}^{1,2} &=&
-i \pi \rho^S_0 \frac{\Delta}{\omega}
-i \pi \rho^S_0 \frac{\Delta}{\omega}
{1 \over {\cal D}_{\rm AF}}
\left\{ -f(x_a) -f(x_b) -f(x_{c,\uparrow})
-f(x_{c,\downarrow})
+2 f(x_a) f(x_{c,\uparrow})
+2 f(x_b) f(x_{c,\downarrow})\right.\\\nonb
&+& f(x_a) f(x_b)
+ f(x_a) f(x_{c,\downarrow})
+ f(x_b) f(x_{c,\uparrow})
+ f(x_{c,\uparrow}) f(x_{c,\downarrow})
- f(x_a) f(x_b) f(x_{c,\downarrow})\\\nonb
&-& \left. f(x_a) f(x_b) f(x_{c,\uparrow})
-  f(x_b) f(x_{c,\uparrow}) f(x_{c,\downarrow})
- f(x_a) f(x_{c,\uparrow}) f(x_{c,\downarrow})
- f(x_a) f(x_b) f(x_{c,\uparrow}) f(x_{c,\downarrow})
\right\}
,
\end{eqnarray}
with
$$
{\cal D}_{\rm AF} = 
1-f(x_b) f(x_{c,\downarrow})
-f(x_a) f(x_{c,\uparrow})
+ f(x_a) f(x_b) f(x_{c,\uparrow})
f(x_{c,\downarrow})
.
$$
The variation of $\Delta^F_{\alpha,\alpha}-
\Delta^{AF}_{\alpha,\alpha}$ as a function of
the transparency of the contact with electrode $c$
is shown on
Fig.~\ref{fig:appli4}. It is visible on
this figure that $\Delta_{\rm F} > \Delta_{\rm AF}$
for all values of the transparency $f(x_c)$
with interface $c$.

%%%%%%%%%%%%%%%%% FIGURE %%%%%%%%%%%%%%%%%%%%%%%%%
%%%%%%%%%%%%%%%%%%%%%%%%%%%%%%%%%%%%%%%%%%%%%%%%%%

%%%%%%%%%%%%%%%%% FIGURE %%%%%%%%%%%%%%%%%%%%%%%%%
\begin{figure}[thb]
\centerline{\fig{8cm}{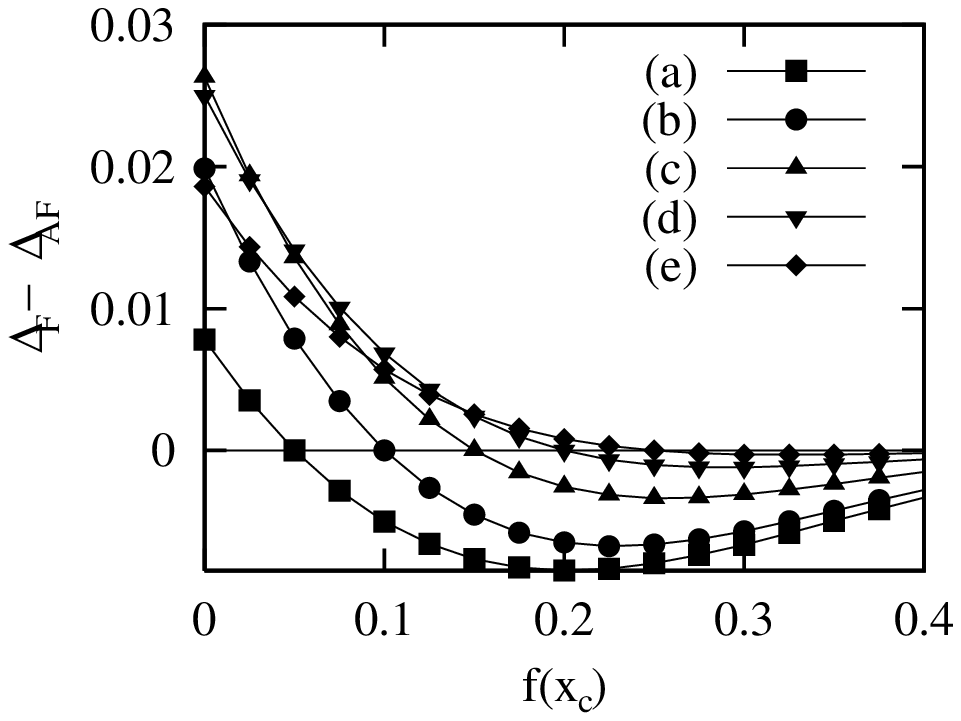}} 
\medskip
\caption{Variation of $\Delta^F_{\alpha,\alpha}
-\Delta^{AF}_{\alpha,\alpha}$ as a function of
the transparency of the contact with the
spin-down ferromagnetic electrode $f(x_{c,\downarrow})$ 
($f(x_{c,\uparrow})=0$) 
for different
values of the transparency of the contacts
with the ferromagnets $f(x_a)=f(x_b)$.
The different curves correspond to
(a) $f(x_a)=f(x_b)=0.05$, (b)
$f(x_a)=f(x_b)=0.1$, (c) $f(x_a)=f(x_b)=0.15$,
(d) $f(x_a)=f(x_b)=0.2$, (e) $f(x_a)=f(x_b)=0.25$.
$\Delta_{\rm F} - \Delta_{\rm AF}$ changes sign as
the transparency of the contact with electrode
$c$ increases.
The microscopic
interaction $U$ is such that the \SOP
of the isolated superconductor is
$\Delta_{\rm bulk}=1$ and the bandwidth of
the superconductor is $D=100$.
} 
\label{fig:appli5}
\end{figure}
%%%%%%%%%%%%%%%%%%%%%%%%%%%%%%%%%%%%%%%%%%%%%%%%%%

We also evaluated $\Delta_{\rm F}-\Delta_{\rm AF}$
in the case
where electrode $c$ is a half-metal ferromagnet
with a spin-down orientation.
We obtain a change of sign in
$\Delta_{\alpha,\alpha}^F
-\Delta_{\alpha,\alpha}^{AF}$ as the transparency
of the interface with the ferromagnetic 
electrode $c$ is increased (see Fig.~\ref{fig:appli5}).
This change of sign can be described by expanding 
the Gorkov functions to second order in 
$f(x_a)$ and $f(x_b)$:
\begin{eqnarray}
G_{\alpha,\alpha}^{1,2,{\rm Ferro}} &=& 
-i \pi \rho^S_0 \frac{\Delta}{\omega}
-i \pi \rho^S_0 \frac{\Delta}{\omega}
\left[ -f(x_a) -f(x_b) -f(x_{c,\uparrow})
-f(x_{c,\downarrow})
+2 f(x_a) f(x_b)
+2 f(x_a) f(x_{c,\uparrow})\right.\\
&+& \left. 2 f(x_b) f(x_{c,\uparrow})
\nonb
+ f(x_a) f(x_{c,\downarrow})
+ f(x_b) f(x_{c,\downarrow})
+ f(x_{c,\uparrow}) f(x_{c,\downarrow})
\right]\\
G_{\alpha,\alpha}^{1,2,{\rm Antiferro}} &=& 
-i \pi \rho^S_0 \frac{\Delta}{\omega}
-i \pi \rho^S_0 \frac{\Delta}{\omega}
\left[ -f(x_a) -f(x_b) -f(x_{c,\uparrow})
-f(x_{c,\downarrow})
+2 f(x_a) f(x_{c,\uparrow})
+2 f(x_b) f(x_{c,\downarrow}) \right.\\
&+& \left.f(x_a) f(x_b)
\nonb
+ f(x_a) f(x_{c,\downarrow})
+ f(x_b) f(x_{c,\uparrow})
+ f(x_{c,\uparrow}) f(x_{c,\downarrow})
\right]
,
\end{eqnarray}
from what we deduce that the difference
between the ferromagnetic and antiferromagnetic
superconducting order parameters changes sign if
$f(x_{c,\downarrow})=f(x_a) + f(x_{c,\uparrow})$,
in agreement with Fig.~\ref{fig:appli5}.
%From this relation we deduce that there is
%no sign change in $\Delta_{\alpha,\alpha}^F
%-\Delta_{\alpha,\alpha}^{AF}$ if
%electrode $c$ has a spin-up orientation.

\subsection{Three node circuit model}
\label{sec:pheno}

The presence or absence of sign changes in
$\Delta_{\alpha,\alpha}^F
-\Delta_{\alpha,\alpha}^{AF}$ discussed in
section~\ref{sec:sign-local}
can also be obtained in the framework
of a phenomenological circuit model~\cite{Melin}.
In this model the \SOP is given by
$$
\Delta = D \exp{\left[ - \frac{1}{U \rho_N}
\left(1+\pi \rho_N \Gamma_\uparrow \right)
\left( 1 + \pi \rho_N \Gamma_\downarrow \right)
\right]}
,
$$
where $\rho_N$ is the density of states in the
normal state, and
where $\Gamma_\sigma$
is the total spectral line-width of spin-$\sigma$
electrons, obtained as the sum of the spectral
line-widths associated to each electrode~\cite{Melin}:
$$
\Gamma_\sigma = \sum_k |t_{S,\alpha_k}|^2 \rho_{k,\sigma}
,
$$
where $t_{S,\alpha_k}$ is the tunnel amplitude connecting
the superconductor and the ferromagnetic electrode $\alpha_k$.

Electrodes $a$ and $b$
are supposed to have the same spin polarization.
We denote by $\gamma=|t_{S,a}|^2 \rho_{a,\uparrow}$
the spectral line-width
associated to majority-spin electrons in electrode $a$
and by $\lambda \gamma=|t_{S,a}|^2 \rho_{a,\downarrow}$
the spectral line-width
associated to minority-spin electrons in electrode $a$. The
spin polarization of the ferromagnetic electrodes
is $P=(1-\lambda)/(1+\lambda)$.

We first suppose that electrode $c$ is
a ferromagnet having a spin-down orientation
with a spectral line-width $\gamma_0$ associated to
majority-spin electrons and a spectral
line-width $\lambda \gamma_0$ associated to
minority-spin electrons. If the
ferromagnets $a$ and $b$ have a parallel
spin orientation the total spectral
line-widths are given by
$\Gamma_\uparrow=2 \gamma+\lambda\gamma_0$ and
$\Gamma_\downarrow =2 \lambda \gamma+\gamma_0$.  If the
ferromagnets $a$ and $b$ have an antiparallel
spin orientation the total spectral
line-widths are given by
$\Gamma_\uparrow=(1+\lambda)\gamma+\lambda \gamma_0$
and $\Gamma_\downarrow
=(1+\lambda)\gamma + \gamma_0$. We obtain
$$
\frac{\Delta_{\rm F}}{\Delta_{\rm AF}}
= \exp{\left[ \frac{\pi^2 \rho_N}{U}
(1-\lambda)^2
(\gamma^2 - \gamma_0^2) \right]}
,
$$
from what we deduce
that $\Delta_{\rm F} > \Delta_{\rm AF}$ if $\gamma_0<\gamma$
and $\Delta_{\rm F} < \Delta_{\rm AF}$ if $\gamma_0>\gamma$,
in agreement with the microscopic model discussed in
section~\ref{sec:local-4channel}.
A similar calculation in the case where electrode
$c$ is a spin-up ferromagnet shows that there
is no sign change in $\Delta_{\rm F} -
\Delta_{\rm AF}$.

The presence of a sign change in $\Delta_{\rm F}
-\Delta_{\rm AF}$ as $\gamma$ increases was
anticipated in section~\ref{sec:physical}.
The value $\gamma=\gamma_0$ at which the change of
sign takes place 
can be understood from the following 
symmetry argument. In the ferromagnetic alignment
of electrodes $a$ and $b$, electrodes $a$ and $b$
have a spin-up orientation and electrode $c$ has
a spin-down orientation. In the antiferromagnetic
alignment electrode $a$ has a spin-up orientation
and electrodes $b$ and $c$ have a spin-down orientation.
Therefore if the three contacts have the same transparency
(namely if $\gamma=\gamma_0$) we see that the case of
a ferromagnetic alignment of electrode $a$ and $b$
can be deduced from the case of an antiferromagnetic
alignment just by reversing the spin of the
three ferromagnetic electrodes. As a consequence
we obtain $\Delta_{\rm F}-\Delta_{\rm AF}=0$
if $\gamma=\gamma_0$ and therefore a sign
change in $\Delta_{\rm F}-\Delta_{\rm AF}$
if $\gamma=\gamma_0$. This argument is also
in agreement with the fact that $\Delta_{\rm F}
-\Delta_{\rm AF}$ changes sign for
$f(x_c)=f(x_a)=f(x_b)$ in the microscopic
model in which three electrodes are connected
to the same site (see Fig.~\ref{fig:appli5}).

We suppose now that electrode $c$ is a normal
metal with a spectral line-width $\gamma_0$
associated to spin-up and spin-down electrons.
We obtain
$$
\frac{\Delta_{\rm F}}{\Delta_{\rm AF}}
= \exp{\left[ \frac{\pi^2 \rho_N}{U} (1-\lambda)^2
\gamma^2 \right]}
$$
which leads to $\Delta_{\rm F}>\Delta_{\rm AF}$
for all values of $\gamma_0$, in agreement
with section~\ref{sec:local-4channel}.

%%%%%%%%%%%%%%%%% FIGURE %%%%%%%%%%%%%%%%%%%%%%%%%
\begin{figure}[thb]
\centerline{\fig{8cm}{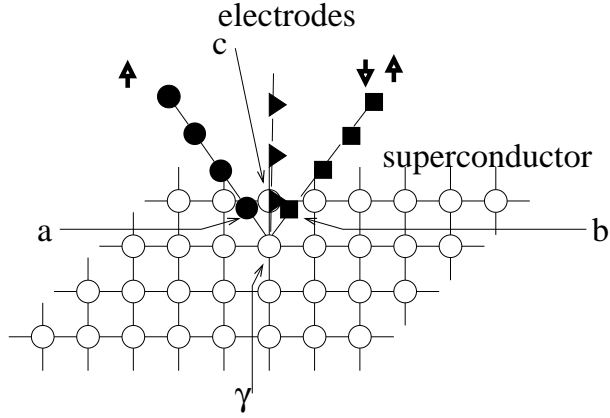}} 
\medskip
\caption{The geometry used in the numerical simulation
in which three one-dimensional
electrodes are connected to a two-dimensional superconductor.
Electrode ending at site
``a'' is ferromagnetic and
is represented by filled circles.
Electrode ending at site
``b'' is ferromagnetic and
is represented by filled squares.
Electrode ending at site
``c'' is a normal metal and is
represented by filled triangles.
}
\label{fig:schema-3elec}
\end{figure}
%%%%%%%%%%%%%%%%%%%%%%%%%%%%%%%%%%%%%%%%%%%%%%%%%%

\subsection{Diagonalizations of the Bogoliubov-de Gennes
Hamiltonian}

We consider exact diagonalizations of the Bogoliubov-de Gennes
Hamiltonian in a situation where a two-dimensional
superconducting system is
connected to two one-dimensional ferromagnetic electrodes
and to a one-dimensional normal metal electrode (see
Fig.~\ref{fig:schema-3elec}).
The proximity effect in the normal metal
electrode can be controlled by the spin orientation
of the two ferromagnetic electrodes in the sense that
the LDOS in the normal
metal electrode depends on the relative
spin orientation of the ferromagnetic electrodes.

\begin{figure}[thb]
\centerline{\fig{6cm}{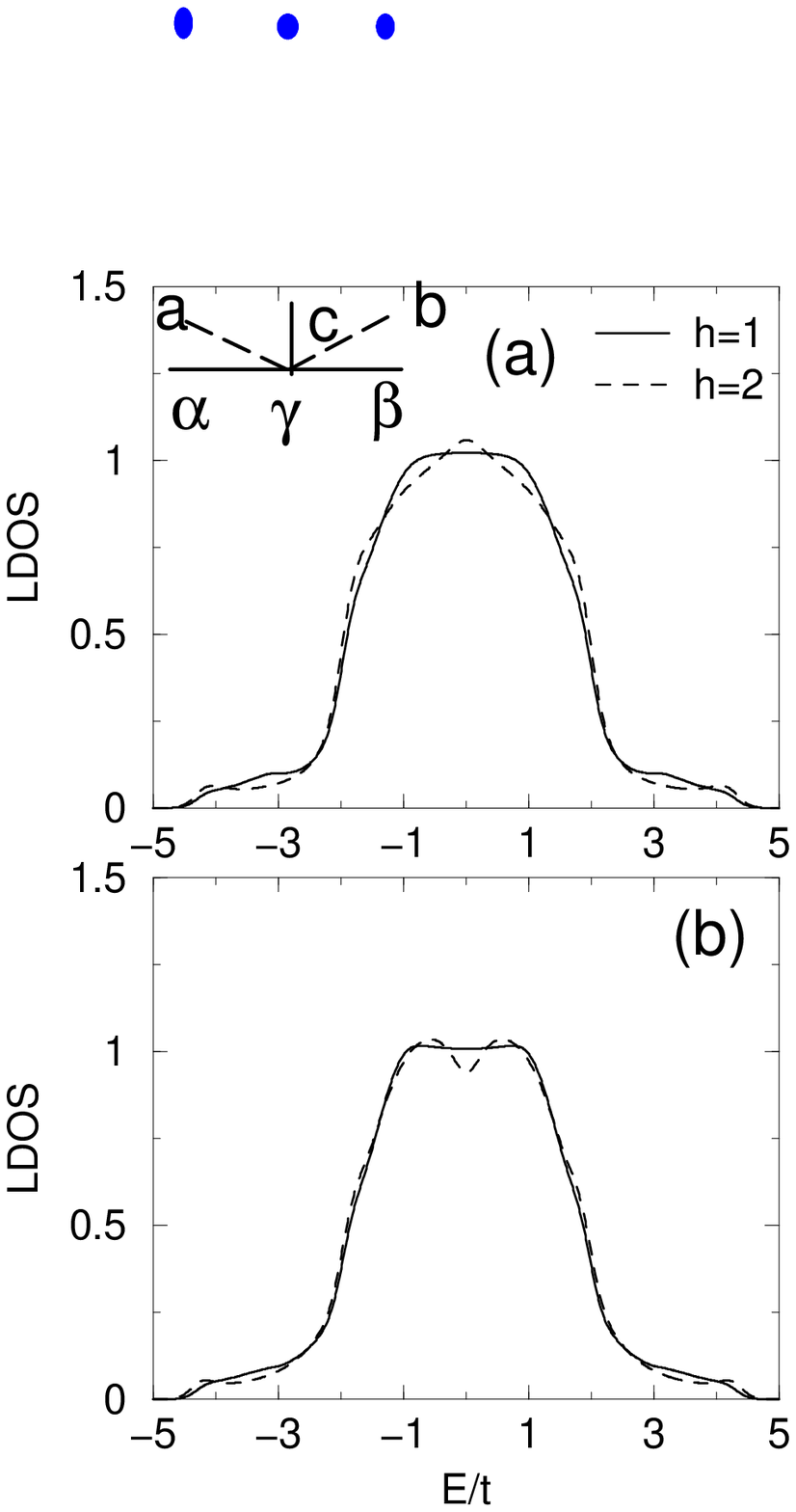}} 
\medskip
\caption{
(a) The LDOS
as a function of $E/t$ at the interface 
of a multiterminal junction made of two 
one-dimensional ferromagnetic electrodes and a
one-dimensional normal metal 
electrode that are connected on top of a two dimensional 
superconducting system. The LDOS is calculated at site $c$
in the normal metal electrode. The upper panel corresponds
to the antiferromagnetic alignment and the lower panel
corresponds to the ferromagnetic alignment. 
}
\label{3dfac.fig}
\end{figure}

The LDOS at site $c$ is shown on
Figs.~\ref{3dfac.fig}(a) and~\ref{3dfac.fig}(b)
in the cases of antiparallel and parallel spin
orientation of electrodes~$a$ and ~$b$.
It is visible that for a sufficiently large exchange
field the LDOS in the normal metal electrode depends
on the relative spin orientation of the ferromagnetic
electrodes. As a consequence it may be possible to
control the proximity effect in the normal metal by
changing the spin orientation of the ferromagnetic
electrodes. This may be of interest in view of 
STM experiments similar to Ref.~\cite{Courtois}.

%%%%%%%%%%%%%%%%% FIGURE %%%%%%%%%%%%%%%%%%%%%%%%%
%%%%%%%%%%%%%%%%%%%%%%%%%%%%%%%%%%%%%%%%%%%%%%%%%%

\section{Friedel oscillations in the Gorkov function}
\label{sec:Friedel-Gorkov}
Sections~\ref{sec:FSF} and~\ref{sec:multiterminal} were
devoted to the analysis of ``local'' models in which several
electrodes are connected to the same site in the
superconductor. Now we consider 
the same problem but with
a finite distance between the ferromagnetic electrodes.
As discussed in section~\ref{sec:physical}
we suppose that the width of the superconductor is small
enough so that pair-breaking effects due to the finite
exchange field can be disregarded.
We consider the geometry on Fig.~\ref{fig:schema1} in which
a superconductor is in contact with two
ferromagnetic electrodes~$a$ and~$a'$. Compared to 
Fig.~\ref{fig:schema1} we use the notation $\alpha'$ and
$a'$ rather than $\beta$ and $b$ for the second electrode.
We keep the notation $\beta$ for an arbitrary site in
the superconductor at which we calculate the \SOP.
We suppose that the two electrodes $a$ and $a'$ are at a distance
much larger than the Fermi wave-length.

\subsection{Form of the Gorkov function}

Electrodes~$a$ and~$a'$
are characterized by the microscopic propagators
$\hat{g}_{a,a}$ and $\hat{g}_{a',a'}$. To order~$t^4$ the
Green's function at site~$\beta$ is given by
\begin{eqnarray}
\label{eq:terme1}
\hat{G}_{\beta,\beta} &=& \hat{g}_{\beta,\beta}
+ \hat{g}_{\beta,\alpha} \hat{t}_{\alpha,a}
\hat{g}_{a,a}
\hat{t}_{a,\alpha} \hat{g}_{\alpha,\beta}
+ \hat{g}_{\beta,\alpha'} \hat{t}_{\alpha',a'}
\hat{g}_{a',a'} \hat{t}_{a',\alpha'} \hat{g}_{\alpha',\beta}\\
\label{eq:terme2}
&+&
\hat{g}_{\beta,\alpha} \hat{t}_{\alpha,a} \hat{g}_{a,a} 
\hat{t}_{a,\alpha}
\hat{g}_{\alpha,\alpha} \hat{t}_{\alpha,a} \hat{g}_{a,a}
\hat{t}_{a,\alpha} \hat{g}_{\alpha,\beta}\\
\label{eq:terme3}
&+& 
\hat{g}_{\beta,\alpha} \hat{t}_{\alpha,a} \hat{g}_{a,a} 
\hat{t}_{a,\alpha}
\hat{g}_{\alpha,\alpha'} \hat{t}_{\alpha',a'} \hat{g}_{a',a'}
\hat{t}_{a',\alpha'} \hat{g}_{\alpha',\beta}\\
\label{eq:terme4}
&+& 
\hat{g}_{\beta,\alpha'} \hat{t}_{\alpha',a'} \hat{g}_{a',a'} 
\hat{t}_{a',\alpha'}
\hat{g}_{\alpha',\alpha} \hat{t}_{\alpha,a} \hat{g}_{a,a}
\hat{t}_{a,\alpha} \hat{g}_{\alpha,\beta}\\
\label{eq:terme5}
&+& 
\hat{g}_{\beta,\alpha'} \hat{t}_{\alpha',a'} \hat{g}_{a',a'} 
\hat{t}_{a',\alpha'}
\hat{g}_{\alpha',\alpha'} \hat{t}_{\alpha',a'} \hat{g}_{a',a'}
\hat{t}_{a',\alpha'} \hat{g}_{\alpha',\beta}
.
\end{eqnarray}
To lowest order the information about the exchange field
in the superconductor is contained in the ``11'' and ``22''
components of the
two terms in Eq.~(\ref{eq:terme1}). The ``12'' component of
the two terms in Eq.~(\ref{eq:terme1}) contains the information
about the effect of a single interface on the Gorkov function
at order $t^2$.
The two terms in Eqs.~(\ref{eq:terme2}) and~(\ref{eq:terme5})
describe a single interface to order $t^4$ and are not directly
relevant to our discussion since they are just a small correction
of the terms in Eq.~(\ref{eq:terme1}) obtained at order $t^2$.
The ``12'' component of the
two terms~(\ref{eq:terme3}) and~(\ref{eq:terme4}) describe
the modification of the Gorkov function due to processes
coupling the two interfaces.
Each of the
two terms~(\ref{eq:terme3}) and~(\ref{eq:terme4})
gives rise to three terms in the Gorkov function
when the summation over spin indexes is made explicit:
\begin{eqnarray}
\label{eq:AA}
(\ref{eq:terme3}) &=& 
\hat{g}_{\beta,\alpha}^{1,2}
t_{\alpha,a}
\hat{g}_{a,a}^{2,2} 
t_{a,\alpha}
\hat{g}_{\alpha,\alpha'}^{2,2}
t_{\alpha',a'}
\hat{g}_{a',a'}^{2,2}
t_{a',\alpha'} \hat{g}_{\alpha',\beta}^{2,2}\\
&+& 
\hat{g}_{\beta,\alpha}^{1,1}
t_{\alpha,a}
\hat{g}_{a,a}^{1,1} 
t_{a,\alpha}
\hat{g}_{\alpha,\alpha'}^{1,2}
t_{\alpha',a'}
\hat{g}_{a',a'}^{2,2}
t_{a',\alpha'} \hat{g}_{\alpha',\beta}^{2,2}\\
&+& 
\hat{g}_{\beta,\alpha}^{1,1}
t_{\alpha,a}
\hat{g}_{a,a}^{1,1} 
t_{a,\alpha}
\hat{g}_{\alpha,\alpha'}^{1,1}
t_{\alpha',a'}
\hat{g}_{a',a'}^{1,1}
t_{a',\alpha'} \hat{g}_{\alpha',\beta}^{1,2}\\
(\ref{eq:terme4})&=&
\hat{g}_{\beta,\alpha'}^{1,2}
t_{\alpha',a'}
\hat{g}_{a',a'}^{2,2}
t_{a',\alpha'}
\hat{g}_{\alpha',\alpha}^{2,2}
t_{\alpha,a}
\hat{g}_{a,a}^{2,2}
t_{a,\alpha}
\hat{g}_{\alpha,\beta}^{2,2}\\
&+& 
\hat{g}_{\beta,\alpha'}^{1,1}
t_{\alpha',a'}
\hat{g}_{a',a'}^{1,1}
t_{a',\alpha'}
\hat{g}_{\alpha',\alpha}^{1,2}
t_{\alpha,a}
\hat{g}_{a,a}^{2,2}
t_{a,\alpha}
\hat{g}_{\alpha,\beta}^{2,2}\\
&+& 
\hat{g}_{\beta,\alpha'_k}^{1,1}
t_{\alpha',a'}
\hat{g}_{a',a'}^{1,1}
t_{a',\alpha'}
\hat{g}_{\alpha',\alpha}^{1,1}
t_{\alpha,a}
\hat{g}_{a,a}^{1,1}
t_{a,\alpha}
\hat{g}_{\alpha,\beta}^{1,2}
\label{eq:BB}
.
\end{eqnarray}

\subsection{Friedel oscillations in the Gorkov function}
Now we use a single channel model to discuss Friedel oscillations
in the Gorkov function.
The terms of order $t^2$ are discussed in section~\ref{sec:t2}
and the terms of order $t^4$ are discussed in section~\ref{sec:t4}.
%%%%%%%%%%%%%%%%% FIGURE %%%%%%%%%%%%%%%%%%%%%%%%%
\begin{figure}[thb]
\centerline{\fig{8cm}{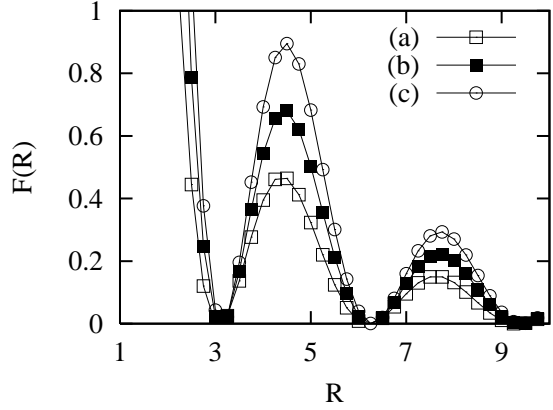}} 
\medskip
\caption{
Variation of $F(R)$ defined by Eq.~(\ref{eq:F(R)}).
We used the parameters $k_F=1$, $\Delta=1$,
$D=10^5$ and $v_F=D/k_F$. The spin-up density of
states is $\rho_{a,\uparrow}=1$. The spin-down
density of states is $\rho_{a,\downarrow}=0$ (a),
$\rho_{a,\downarrow}=0.5$ (b) and
$\rho_{a,\downarrow}=1$ (c).
}
\label{fig:F(R)}
\end{figure}
%%%%%%%%%%%%%%%%%%%%%%%%%%%%%%%%%%%%%%%%%%%%%%%%%%

\subsubsection{Terms of order $t^2$}
\label{sec:t2}
Let us start with the terms of order $t^2$. Since these terms do
not couple the two interfaces we consider a single interface only.
We obtain easily the expression of the Gorkov function
to order $t^2$:
\begin{eqnarray}
G_{\beta,\beta}^{1,2,A} &=& -i \frac{2 m a_0^2}{\hbar^2}
\frac{a_0}{2\pi R_0} \frac{\Delta}{\sqrt{\omega^2-\Delta^2}}\\
&-&\left( \frac{2 m a_0^2}{\hbar^2} \right)^2
\left( \frac{a_0}{2\pi R} \right)^2
\pi t_\alpha^2
\left[ \rho_{a,\uparrow} e^{-i k_F R}
-\rho_{a,\downarrow} e^{i k_F R} \right]
\sin{\left[ k_F R \right]}
\exp{\left[ - \frac{2 i \sqrt{\omega^2-\Delta^2} R}{v_F}
\right]} \frac{\Delta}{\sqrt{\omega^2-\Delta^2}}
.
\end{eqnarray}
If site $\beta$ is equal to site $\alpha$ we use
the regularization $k_F R_0=\pi/2$ discussed in
section~\ref{sec:NF} to obtain the self-consistent
order parameter at the interface:
$$
\Delta_\alpha = 2 D \exp{\left[ -\frac{1}{U}
\left\{ \frac{2 m a_0^2}{\pi \hbar^2} \frac{a_0}{2 \pi R_0}
\left[ 1 - \frac{2 m a_0^2}{\hbar^2} \frac{a_0}{2\pi R_0}
\pi t_\alpha^2 \left( \rho_{a,\uparrow} + \rho_{a,\downarrow}
\right) \right] \right\}^{-1} \right]}
.
$$
As expected the superconducting order parameter at the interface
is reduced because of the proximity effect. 

To discuss the proximity effect in the bulk of the superconductor
we write the pair amplitude under the form
$$
\int_\Delta^D \mbox{Im} \left[ G_{\beta,\beta}^{1,2,A} \right]
d \omega
= - \frac{2 m a_0^2}{\hbar^2} \frac{a_0}{2\pi R_0}
\int_\Delta^D \frac{\Delta}{\sqrt{\omega^2-\Delta^2}} d \omega
+ \left( \frac{2 m a_0^2}{\hbar^2} \right)^2
\left( \frac{a_0}{2\pi} \right)^2 \pi t_\alpha^2
F(R)
,
$$
with
\begin{equation}
\label{eq:F(R)}
F(R) = \frac{ \sin{[k_F R]}}{R^2}
\int_\Delta^D \left[ \rho_{a,\uparrow}
\sin{ \left\{ \left[ k_F +
\frac{2 \sqrt{\omega^2-\Delta^2}}{v_F} R \right] \right\}}
+
\rho_{a,\downarrow}
\sin{ \left\{ \left[ k_F -
\frac{2 \sqrt{\omega^2-\Delta^2}}{v_F} R \right] \right\}} \right]
 \frac{\Delta}{\sqrt{\omega^2-\Delta^2}}
d\omega
.
\end{equation}
The variations of $F(R)$ as a function of $R$ are shown
on Fig.~\ref{fig:F(R)}. It is visible that $F(R)$ oscillates
in space but is always positive which means that
the superconducting
order parameter is reduced in the bulk of the
superconductor and shows Friedel oscillations.

%%%%%%%%%%%%%%%%% FIGURE %%%%%%%%%%%%%%%%%%%%%%%%%
\begin{figure}[thb]
\centerline{\fig{8cm}{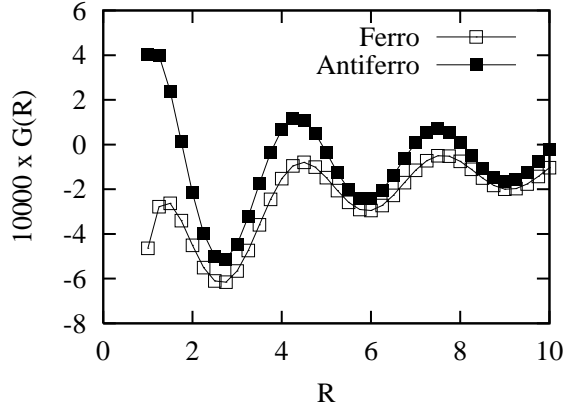}} 
\medskip
\caption{
Variation of $G(R)$ defined by Eq.~(\ref{eq:G(R)}).
We used the parameters $k_F=1$, $\Delta=1$,
$D=10^5$ and $v_F=D/k_F$. The spin-up density of
states is $\rho_{a,\uparrow}=1$. The spin-down
density of states is $\rho_{a,\downarrow}=0.2$.
The distance between the two interfaces is $R_{\alpha,\alpha'}
=100$ and the function $G(R)$ is calculated as a function
of $R=R_{\alpha,\beta} = R_{\alpha,\alpha'}
-R_{\alpha',\beta}$.
}
\label{fig:G(R)}
\end{figure}
%%%%%%%%%%%%%%%%%%%%%%%%%%%%%%%%%%%%%%%%%%%%%%%%%%

\subsubsection{Terms of order $t^4$}
\label{sec:t4}
Now we consider the terms of order $t^4$ that couple the two interfaces,
namely the terms~(\ref{eq:terme3}) and~(\ref{eq:terme4}).
Using Eqs.~(\ref{eq:AA})~-- (\ref{eq:BB}) we obtain
\begin{equation}
\mbox{Im}\left[ (\ref{eq:terme3}) + (\ref{eq:terme4}) \right]
= \frac{1}{2} \pi^2 t_\alpha^2 t_{\alpha'}^2
\left( \frac{2 m a_0^2}{\hbar^2} \right)^3
\left( \frac{a_0}{2\pi} \right)^3 G(R)
,
\end{equation}
with
\begin{eqnarray}
\label{eq:G(R)}
&& G(R) =
\frac{1}{R_{\alpha,\beta} R_{\alpha,\alpha'} R_{\alpha',\beta}}
\times \left\{
2 \rho_{a,\uparrow} \rho_{a',\downarrow}
\int_\Delta^D d\omega \frac{\Delta}{\sqrt{\omega^2-\Delta^2}}
\sin{\left\{ \left[ k_F - \frac{\omega^2-\Delta^2}{v_F}
\right] \left[ R_{\alpha,\beta} + R_{\alpha,\alpha'}
+R_{\alpha',\beta} \right] \right\} }
\right.\\
&+& 2 \rho_{a,\uparrow} \rho_{a',\uparrow}
\int_\Delta^D d \omega \frac{\Delta}{\sqrt{\omega^2-\Delta^2}}
\sin{\left\{ \left[ k_F + \frac{\sqrt{\omega^2-\Delta^2}}{v_F}
\right] \left[ R_{\alpha,\beta} + R_{\alpha,\alpha'}
+R_{\alpha',\beta} \right] \right\} }\\
&+& \left(\rho_{a,\uparrow} + \rho_{a,\downarrow} \right)
\rho_{a',\downarrow} 
\int_\Delta^D d\omega \frac{\Delta}{\sqrt{\omega^2-\Delta^2}}
\sin{ \left\{ \left[ k_F+\frac{\sqrt{\omega^2-\Delta^2}}{v_F}
\right] R_{\alpha,\beta}
-\left[ k_F - \frac{ \sqrt{\omega^2-\Delta^2}}{v_F}
\right] \left[ R_{\alpha,\alpha'} + R_{\alpha',\beta} \right]
\right\} }\\
&+& \left(\rho_{a,\uparrow} + \rho_{a,\downarrow} \right)
\rho_{a',\uparrow} 
\int_\Delta^D d\omega \frac{\Delta}{\sqrt{\omega^2-\Delta^2}}
\sin{ \left\{ \left[ k_F-\frac{\sqrt{\omega^2-\Delta^2}}{v_F}
\right] R_{\alpha,\beta}
-\left[ k_F + \frac{ \sqrt{\omega^2-\Delta^2}}{v_F}
\right] \left[ R_{\alpha,\alpha'} + R_{\alpha',\beta} \right]
\right\} }\\
&+& \rho_{a,\uparrow} 
\left(\rho_{a',\uparrow} + \rho_{a',\downarrow} \right)
\int_\Delta^D d\omega \frac{\Delta}{\sqrt{\omega^2-\Delta^2}}
\sin{ \left\{ -\left[ k_F+\frac{\sqrt{\omega^2-\Delta^2}}{v_F}
\right] \left[R_{\alpha,\beta}+R_{\alpha,\alpha'} \right]
+\left[ k_F - \frac{ \sqrt{\omega^2-\Delta^2}}{v_F}
\right] R_{\alpha',\beta}
\right\} }\\
&+& \left. \rho_{a,\downarrow} 
\left(\rho_{a',\uparrow} + \rho_{a',\downarrow} \right)
\int_\Delta^D d\omega \frac{\Delta}{\sqrt{\omega^2-\Delta^2}}
\sin{ \left\{ -\left[ k_F-\frac{\sqrt{\omega^2-\Delta^2}}{v_F}
\right] \left[R_{\alpha,\beta}+R_{\alpha,\alpha'} \right]
+\left[ k_F + \frac{ \sqrt{\omega^2-\Delta^2}}{v_F}
\right] R_{\alpha',\beta}
\right\} } \right\}
.
\end{eqnarray}
The variations of $G(R)$ are shown on Fig.~\ref{fig:G(R)}.
It is visible that $G(R)$ is larger in the antiferromagnetic
alignment and therefore in the bulk of the superconductor
the \SOP is larger in the ferromagnetic alignment.
This shows that the effect is not a specificity of the
local models analyzed in sections~\ref{sec:FSF}
and~\ref{sec:multiterminal}
in which all electrodes are connected to the same site.
On the contrary the effect occurs also in the models
on Figs.~\ref{fig:schema1}-(b) and~\ref{fig:schema2}-(b)
in which the distance between the ferromagnetic electrodes
is large compared to the Fermi wave-length.

\section{Conclusion}
\label{sec:conclusion}
To conclude we have provided a detailed investigation
of the proximity effect in
multi-connected hybrid structures in which
several electrodes are connected to a superconductor.
We have pointed out the existence of two mechanisms
involved in the determination of the \SOP.
One is related to the existence of
an exchange field in the superconductor that
was first pointed out in Ref.~\cite{deGennes}.
The other mechanism takes place already at a single NS
interface.
We have first reconsidered FSF heterostructures
in which two ferromagnetic electrodes are
connected to a superconductor at a distance smaller
than the superconducting coherence length. We have
found that within a ``local'' model the \SOP
in the ferromagnetic alignment is larger than
the \SOP in the antiferromagnetic alignment,
in agreement with a different model discussed
in Ref.~\cite{Apinyan}. We have shown that
the zero-energy LDOS in the ferromagnetic
alignment is larger than the zero-energy
LDOS in the antiferromagnetic alignment.
In the case where two ferromagnetic and a normal
metal electrode are connected to a superconductor
we have found that the LDOS in the normal metal
depends on the spin orientation in
the ferromagnetic electrodes. 
If the normal
metal is replaced by a ferromagnetic metal
with a spin-down orientation
we have found that $\Delta_{\rm AF} > \Delta_{\rm F}$
for high transparencies and we have provided
two analytical models for this behavior
(an inversion of the $4 \times 4$ Dyson matrix
for half-metal ferromagnets and another approach
based on a circuit model for partially
polarized ferromagnets). We have pointed out that
this behavior could be understood from a simple
rule stating that for this model the increase of
pair correlations in the ferromagnetic electrodes
generates a reduction of the \SOP at the interface.
This behavior is
confirmed by the exact diagonalizations of the
Bogoliubov-de Gennes equations for large values of
the exchange field but not for small values of
the exchange field. 

One explanation to the discrepancy between
the two models is that in the Green's function
approach only the contribution of energies far
above the superconducting gap has been taken
into account. This is legitimate if the
band-width $D$ of the superconductor is much larger
than the superconducting gap
(in real systems one has typically $D/\Delta
\simeq 10^5$). This is justified by the fact that the
integral in Eq.~(\ref{eq:self-con}) is
dominated by the high-energy behavior of the
Gorkov function $G^{+,-,1,2}_{\beta,\beta}
(\omega) \sim 1/\omega$. On the other
hand in the exact diagonalizations of the
Bogoliubov-de Gennes Hamiltonian we used small
system sizes and replaced the right hand side of
Eq.~(\ref{eq:self-con}) by a sum over all energy levels.
It is thus not surprising that
the two procedures (namely the high-energy behavior
for large systems and a sum over all energy levels
for small systems) can lead to different results
in some cases. Another possible source of
discrepancy between the two approaches lies in the
fact that the ferromagnetic electrodes are one-dimensional
and have a finite size in the exact diagonalizations
whereas the ferromagnetic electrodes are described
by the local density of states of a three dimensional
metal
in the Green's function approach.

\section*{Acknowledgement}
The authors acknowledge fruitful discussions
with H. Courtois, D. Feinberg, M. Giroud.

\end{document}